\documentclass[a4paper,fleqn,usenatbib]{mnras}

\usepackage[T1]{fontenc}
\usepackage{ae,aecompl}

\usepackage{graphicx}	
\usepackage{amsmath}	
\usepackage{amssymb}	
\usepackage{subfigure}
\usepackage{multirow}
\usepackage{xcolor}
\hypersetup{draft}

\newcommand{\pcc}{\,{\rm cm}^{-3}}
\newcommand{\gcc}{\,{\rm g \, cm}^{-3}}
\newcommand{\pcs}{\,{\rm cm}^{-2}}

\newcommand{\kel}{\, {\rm K}}

\newcommand{\msun}{\, {\rm M}_\odot}
\newcommand{\nh}{n_{\rm H}}

\newcommand{\pc}{\, {\rm pc}}

\newcommand{\myr}{\, {\rm Myr}}
\newcommand{\kyr}{\, {\rm kyr}}
\newcommand{\ug}{\, {\rm \mu G}}

\newcommand{\kms}{\, {\rm km \, s^{-1}}}

\title[Early COM formation in molecular clouds]{NEATH IV: an early onset of complex organic chemistry in molecular clouds}

\author[Priestley et al.]{
  F. D. Priestley$^1$\thanks{Email: priestleyf@cardiff.ac.uk}, P. C. Clark$^1$, S. E. Ragan$^1$, S. Scibelli$^2$\thanks{Jansky Fellow of the National Radio Astronomy Observatory}, M. T. Cusack$^1$,
  \newauthor S. C. O. Glover$^3$, O. Feh\'{e}r$^1$, L. R. Prole$^4$, R. S. Klessen$^{3,5,6,7}$
\\
$^{1}$School of Physics and Astronomy, Cardiff University, Queen's Buildings, The Parade, Cardiff CF24 3AA, UK \\
$^{2}$National Radio Astronomy Observatory, 520 Edgemont Road, Charlottesville 22903-2475, USA \\
$^{3}$Universit\"{a}t Heidelberg, Zentrum f\"{u}r Astronomie, Institut f\"{u}r Theoretische Astrophysik, Albert-Ueberle-Stra{\ss}e 2, D-69120 Heidelberg, Germany \\
$^{4}$Centre for Astrophysics and Space Science Maynooth, Department of Theoretical Physics, Maynooth University, W23 F2H6 Maynooth, Ireland \\
$^{5}$Universit\"{a}t Heidelberg, Interdisziplin\"{a}res Zentrum f\"{u}r Wissenschaftliches Rechnen, Im Neuenheimer Feld 205, D-69120 Heidelberg, Germany \\
$^{6}$Harvard-Smithsonian Center for Astrophysics, 60 Garden Street, Cambridge, MA 02138, U.S.A. \\
$^{7}$Elizabeth S. and Richard M. Cashin Fellow at the Radcliffe Institute for Advanced Studies at Harvard University, 10 Garden Street, Cambridge, MA 02138, U.S.A.
}

\date{Accepted XXX. Received YYY; in original form ZZZ}

\pubyear{2025}

\begin{document}
\label{firstpage}
\pagerange{\pageref{firstpage}--\pageref{lastpage}}
\maketitle

\begin{abstract}

  Complex organic molecules (COMs) are widely detected in protostellar and protoplanetary systems, where they are thought to have been inherited in large part from earlier evolutionary phases. The chemistry of COMs in these earlier phases, namely starless and prestellar cores, remains poorly understood, as models often struggle to reproduce the observed gas-phase abundances of these species. We simulate the formation of a molecular cloud, and the cores within it, out of the diffuse interstellar medium, and follow the chemical evolution of the cloud material starting from purely-atomic initial conditions. We find that the formation of both gas- and ice-phase COMs precedes the formation of cores as distinct objects, beginning at gas densities of a few $10^3 \pcc$. Much of this COM-enriched material remains at these relatively modest densities for several $\myr$, which may provide a reservoir for accretion onto planet-forming discs in later evolutionary stages. We suggest that models of core and disc chemistry should not ignore the complex dynamical evolution which precedes these structures, even when studying supposedly late-forming molecules such as CH$_3$OH and CH$_3$CN.

\end{abstract}

\section{Introduction}

Complex organic molecules (COMs), in an astrochemical context, are those made up of six or more atoms, including at least one of carbon \citep{herbst2009}. They are widely detected towards protostellar and protoplanetary systems via both emission from gas-phase COMs \citep[e.g.][]{walsh2016,jorgensen2016,ceccarelli2017} and, increasingly, mid-infrared absorption features caused by COMs in icy mantles around dust grains \citep{mcclure2023,rocha2024,nazari2024}. In either form, they are likely to be incorporated into planets forming within these systems \citep{oberg2021}, and represent a potential starting point for the formation of amino acids and other molecules with important biological roles on Earth. There has thus been significant interest in the potential connection between interstellar COMs and the origins of life \citep[e.g.][]{jimenez2020}.

The presence of COMs in protostellar cores is generally understood to result from their formation in ice mantles during the preceeding prestellar phase, which are subsequently evaporated into the gas phase by protostellar heating \citep{viti2004,garrod2006}. This picture struggles to explain the widespread detection of gas-phase COMs in starless and prestellar cores \citep{jimenez2016,scibelli2020,ambrose2021,scibelli2021,jimenez2021,mininni2021,punanova2022,megias2023}, which have no internal heating source. Detections of COMs in diffuse and translucent clouds \citep[e.g.][]{liszt2018} raise similar problems, as the densities should be to low to allow their formation on grain surfaces. Suggested resolutions to these issues include enhanced reactive desorption efficiencies \citep{vasyunin2017,riedel2023}, external irradiation \citep{spezzano2022,jensen2023}, and nondiffusive surface chemistry \citep{jin2020,garrod2022}, but at present no consensus has been reached on the correct explanation. As the chemical makeup of protostellar systems is in large part inherited from the prestellar phase \citep{booth2021,jensen2021b}, this severely limits our ability to understand the delivery of COMs to planets forming within these systems.

The vast majority of modelling efforts regarding COM chemistry in cores neglect the physical evolution of the system entirely, assuming the density (or density profile) remains static for the entire $\sim \myr$ duration of the chemical simulation. Models also typically assume the chemical initial conditions are atomic (with the exception of H$_2$), or representative of a prior `diffuse cloud' evolutionary phase. However, cores form in the dynamic environments of molecular clouds, which can have a significant impact on their chemistry when compared to static models \citep{jensen2021,clement2023,priestley2023a}. In particular, molecular cloud material can undergo multiple transient enhancements to relatively high densities ($> 10^3 \pcc$) before being incorporated into a gravitationally-bound core \citep{priestley2023b}, so the assumption of chemical initial conditions corresponding to atomic or diffuse molecular gas is unlikely to be applicable to real objects.

In this paper, we investigate the origins of COMs starting from the cold neutral phase of the interstellar medium (ISM), where even hydrogen is in atomic form. Using hydrodynamical simulations coupled to a time-dependent gas-grain chemical network \citep{priestley2023a}, we can simultaneously and self-consistently follow both the formation of cores from the diffuse ISM, and the build-up of chemical complexity from pristine initial conditions. We find that the formation of significant quantities of COMs occurs before the formation of the cores themselves as distinct objects. The molecular composition of protostellar systems may be inherited not just from the prestellar phase but from the parent molecular cloud itself, making it essential to consider the full evolutionary history of a system when modelling its chemistry.

\section{Method}

\begin{table}
  \centering
  \caption{Elemental abundances used in the chemical modelling.}
  \begin{tabular}{ccccc}
    \hline
    Element & Abundance & & Element & Abundance \\
    \hline
    C & $1.4 \times 10^{-4}$ & & S & $1.2 \times 10^{-5}$ \\
    N & $7.6 \times 10^{-5}$ & & Si & $1.5 \times 10^{-7}$ \\
    O & $3.2 \times 10^{-4}$ & & Mg & $1.4 \times 10^{-7}$ \\
    \hline
  \end{tabular}
  \label{tab:abun}
\end{table}

\begin{figure*}
  \centering
  \includegraphics[width=0.85\textwidth]{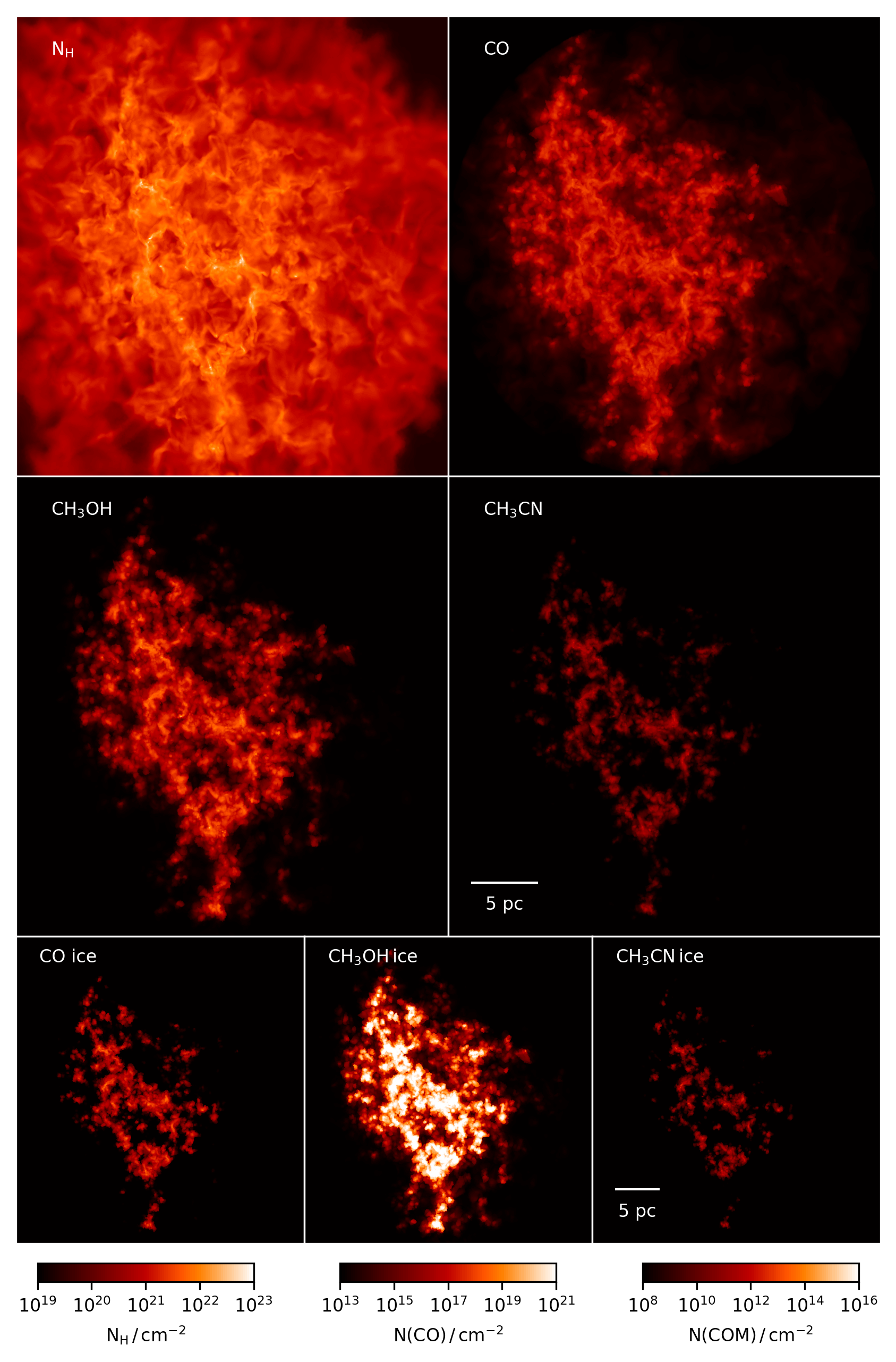}
  \caption{Column density maps at $5.53 \myr$ of total hydrogen nuclei, and of CO, CH$_3$OH and CH$_3$CN in gas and ice phases. The colour bars correspond to total hydrogen nuclei (left) and the gas- and ice-phase columns of CO (centre) and of both COM species (right).}
  \label{fig:coldens}
\end{figure*}

\begin{figure*}
  \centering
  \includegraphics[width=0.32\textwidth]{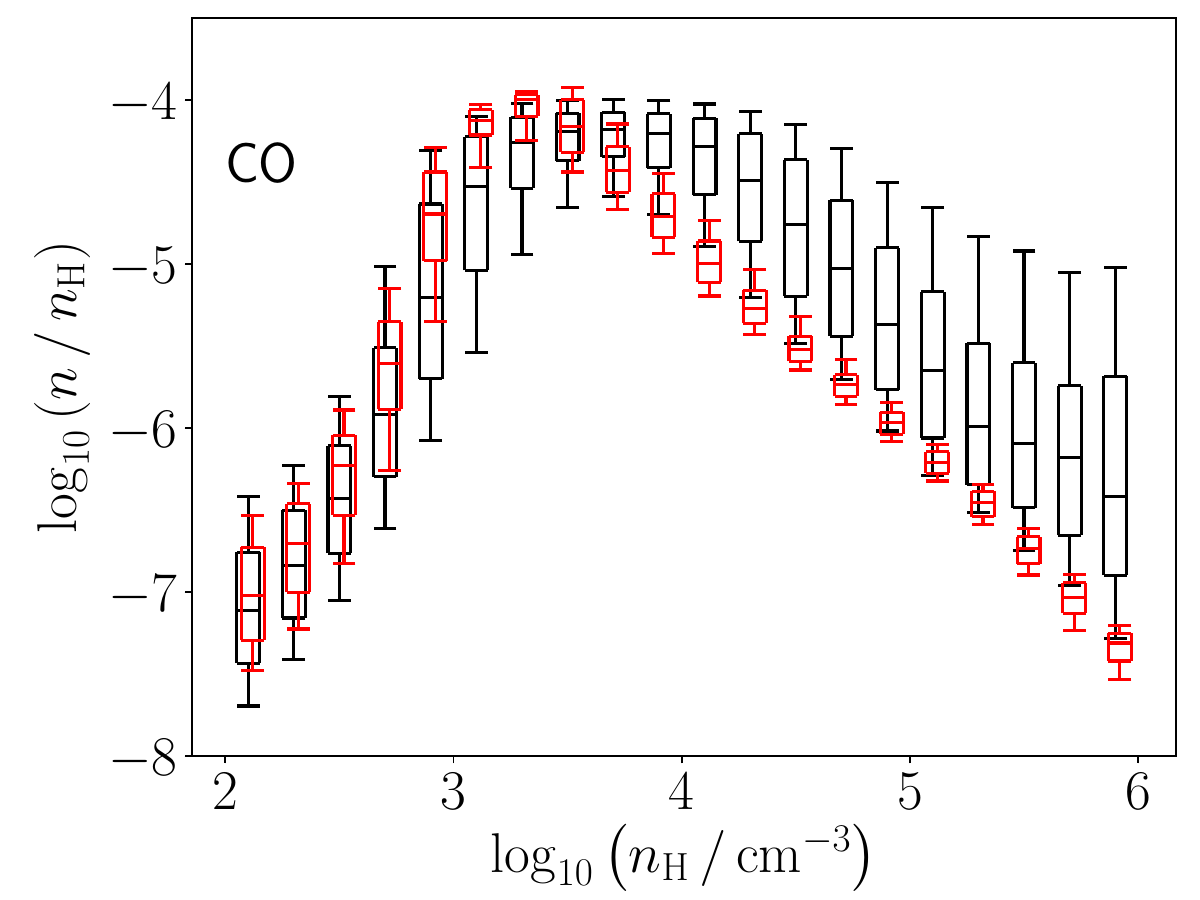}
  \includegraphics[width=0.32\textwidth]{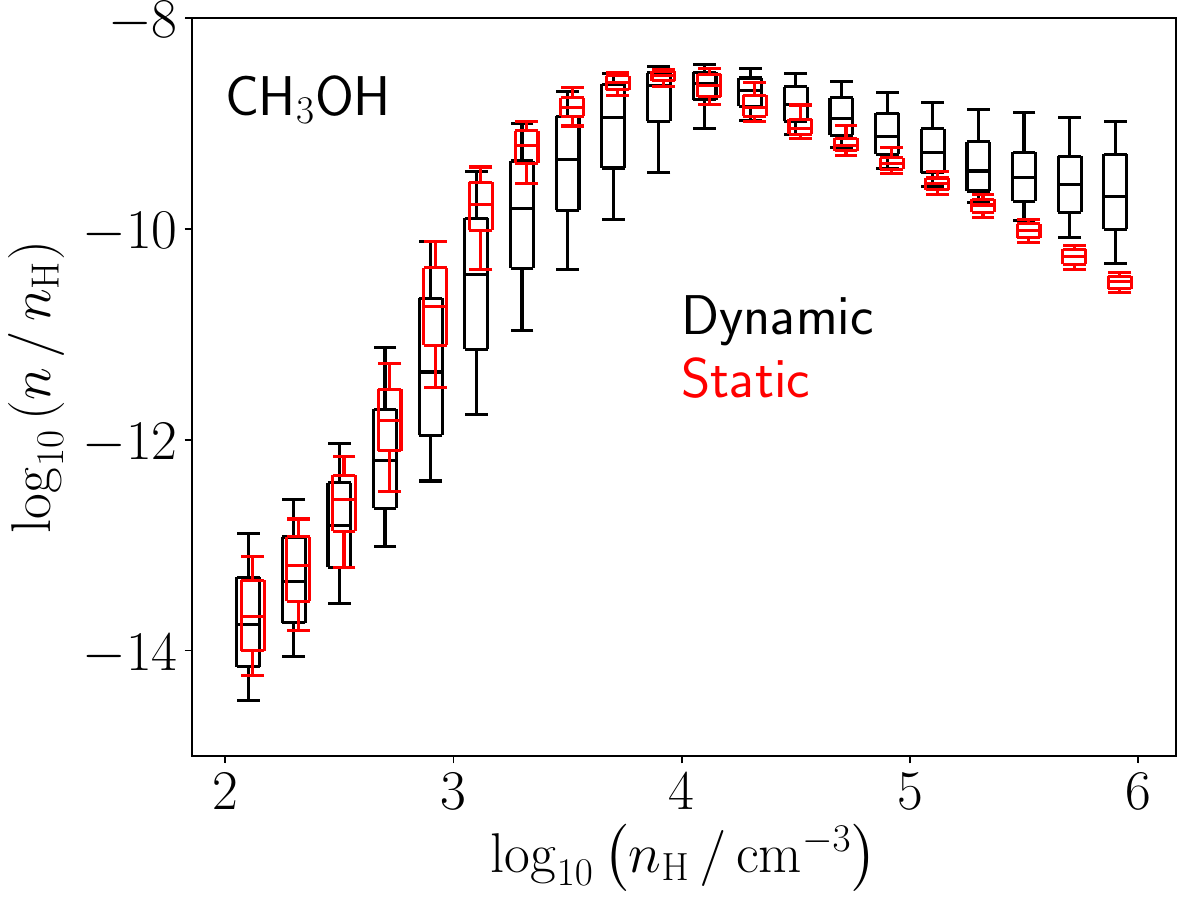}
  \includegraphics[width=0.32\textwidth]{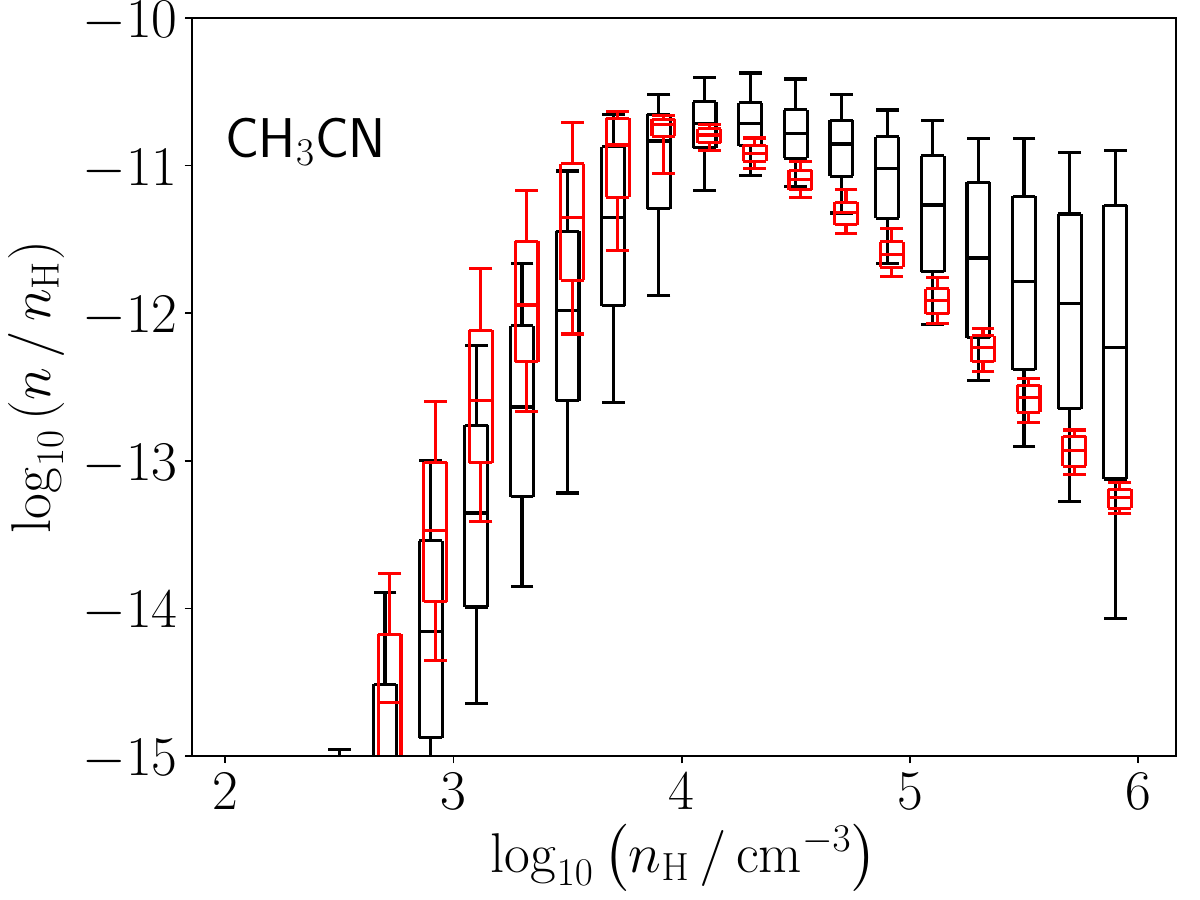}\\
  \includegraphics[width=0.32\textwidth]{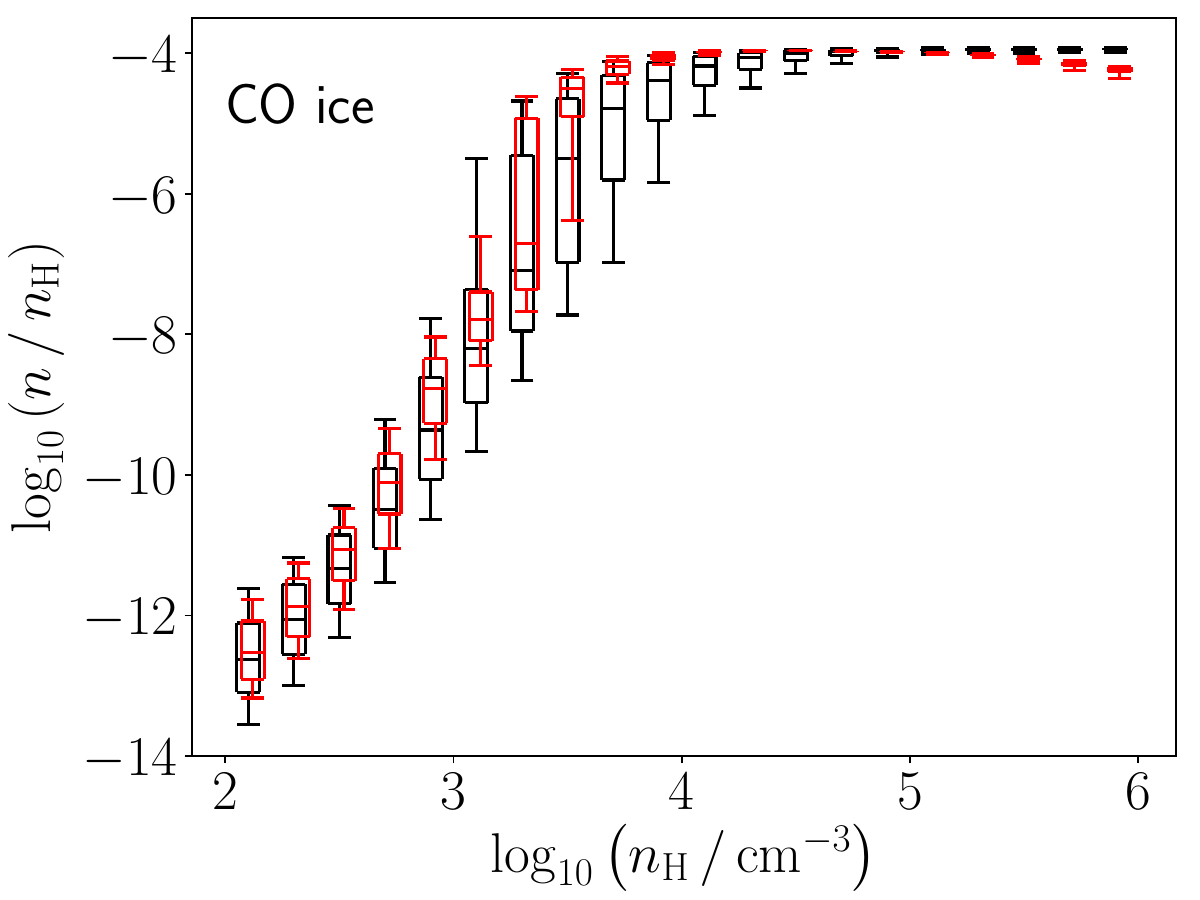}
  \includegraphics[width=0.32\textwidth]{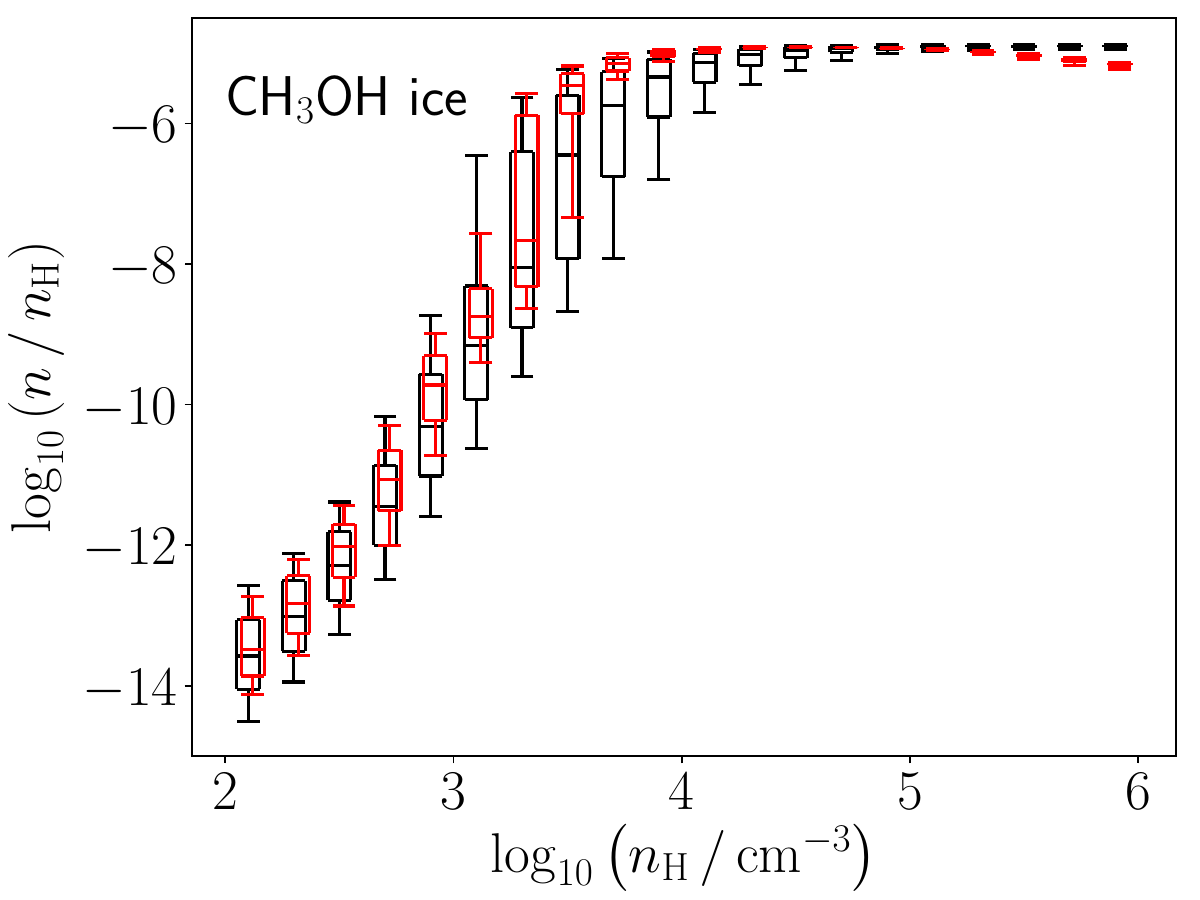}
  \includegraphics[width=0.32\textwidth]{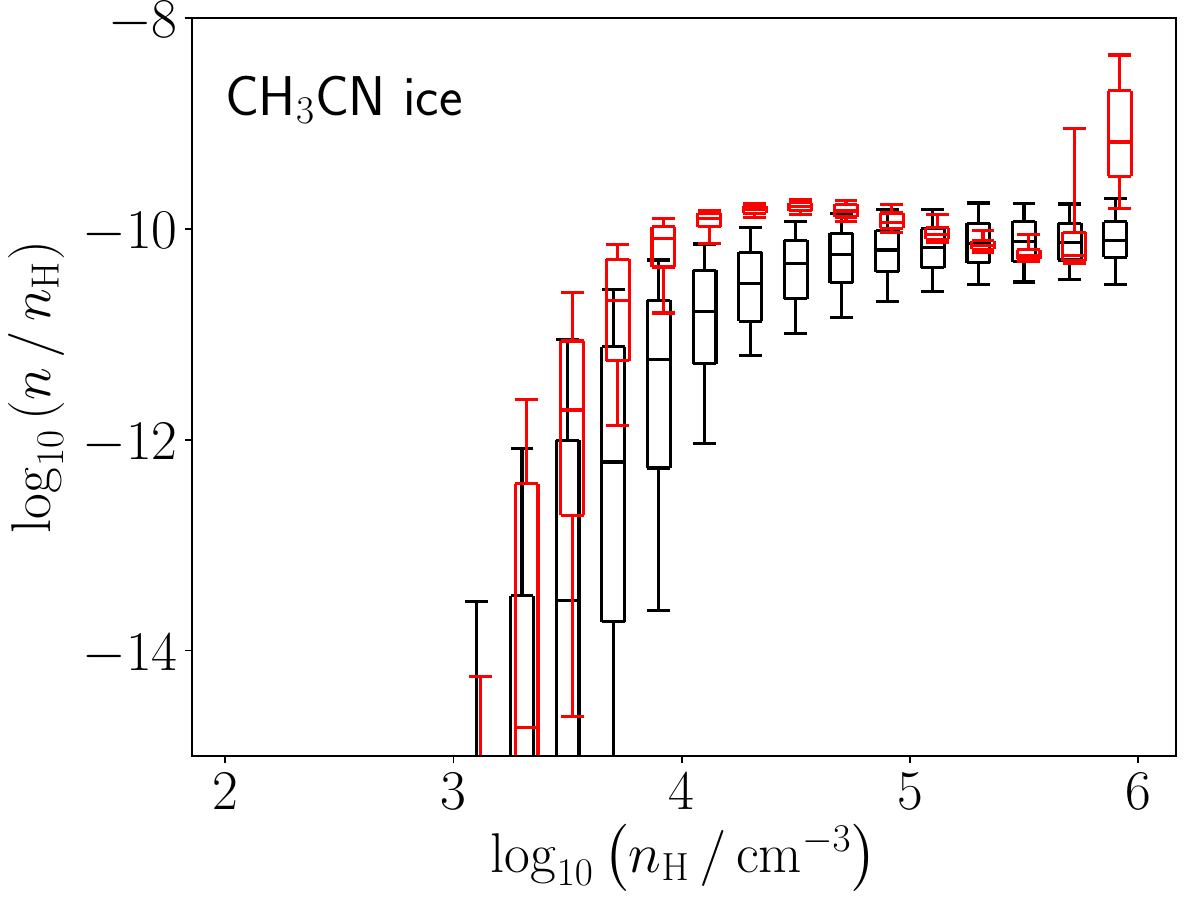}
  \caption{Abundances of gas-phase (top row) and ice-phase (bottom row) CO (left), CH$_3$OH (centre) and CH$_3$CN (right) as a function of gas density at $5.53 \myr$, for chemical models utilising the full tracer histories (black) or holding physical properties constant at their final values (red). Boxes show median values and 25th/75th percentiles, whiskers the 10th/90th percentiles.}
  \label{fig:volabun}
\end{figure*}

\begin{figure*}
  \centering
  \includegraphics[width=0.32\textwidth]{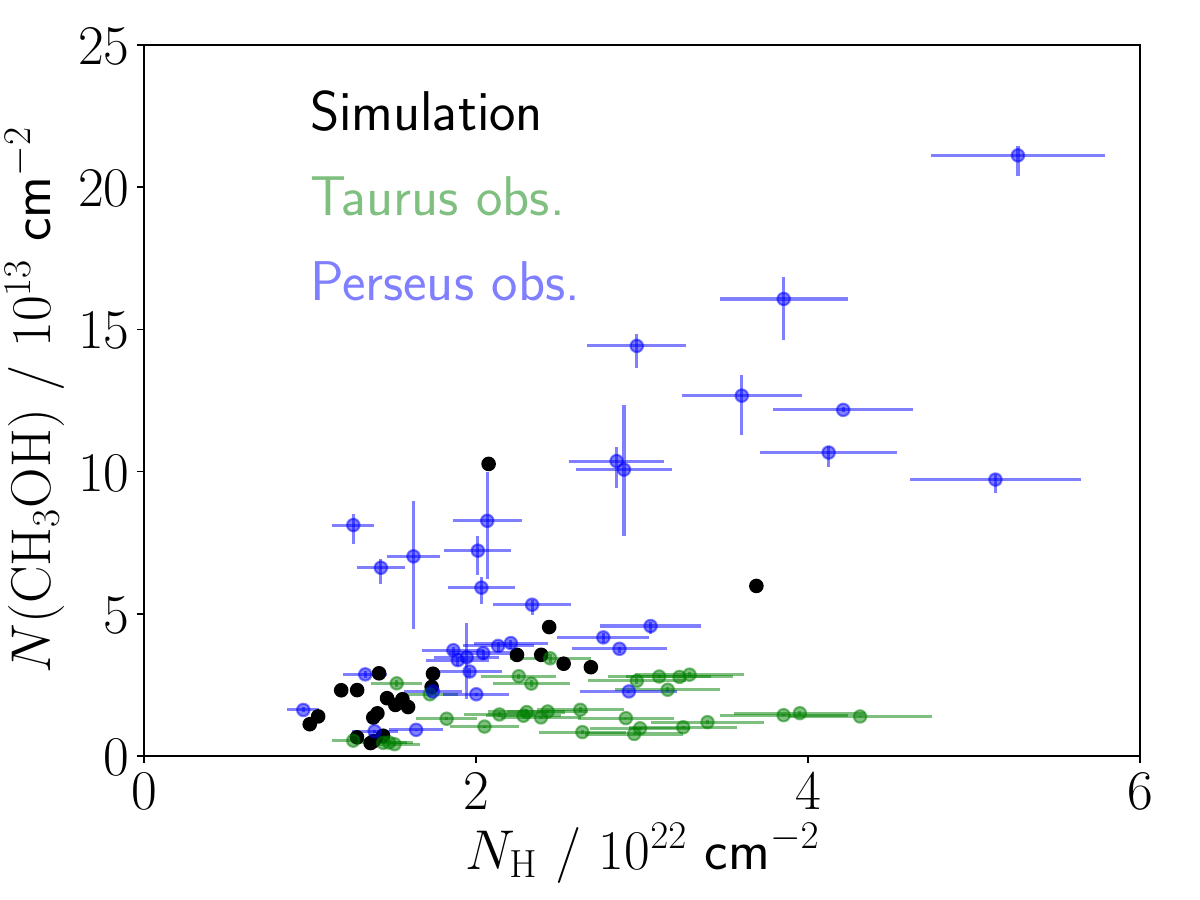}
  \includegraphics[width=0.32\textwidth]{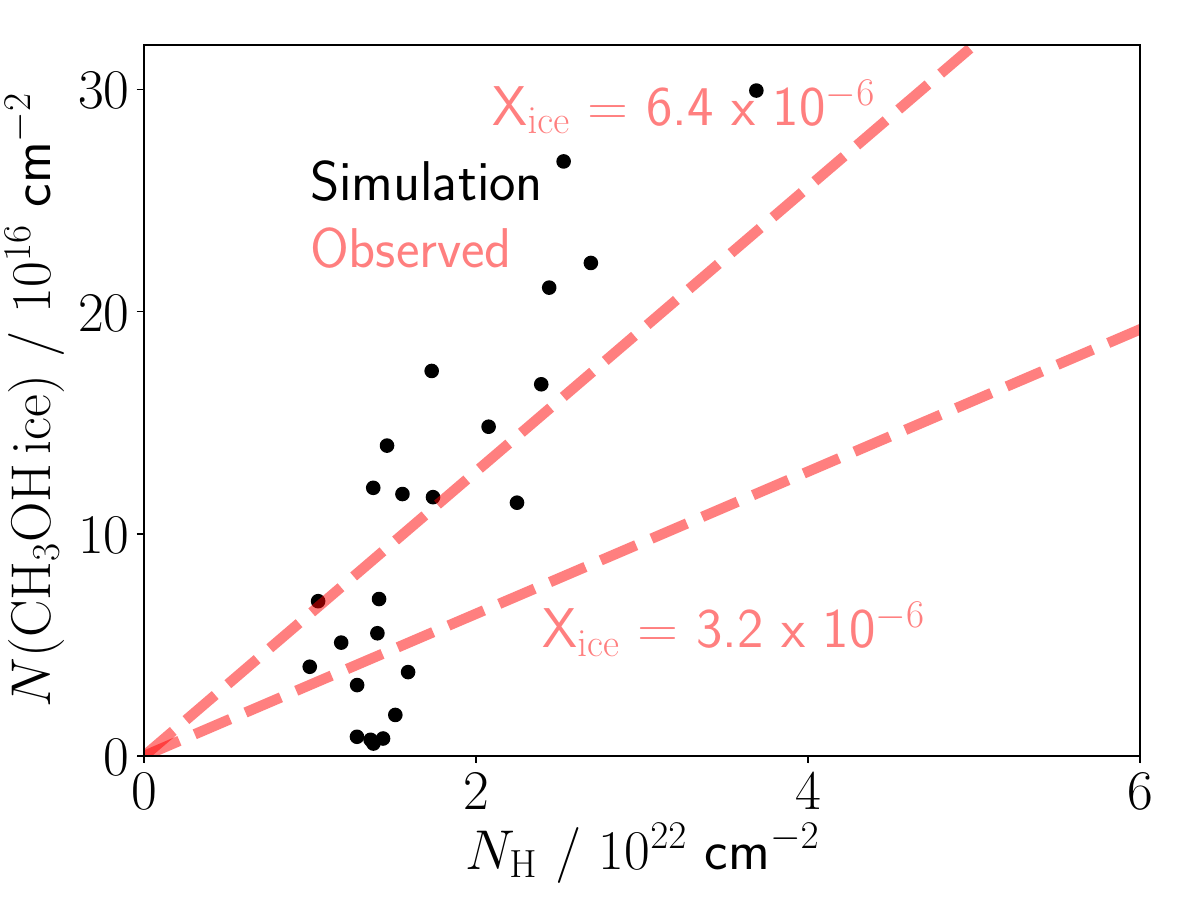}
  \includegraphics[width=0.32\textwidth]{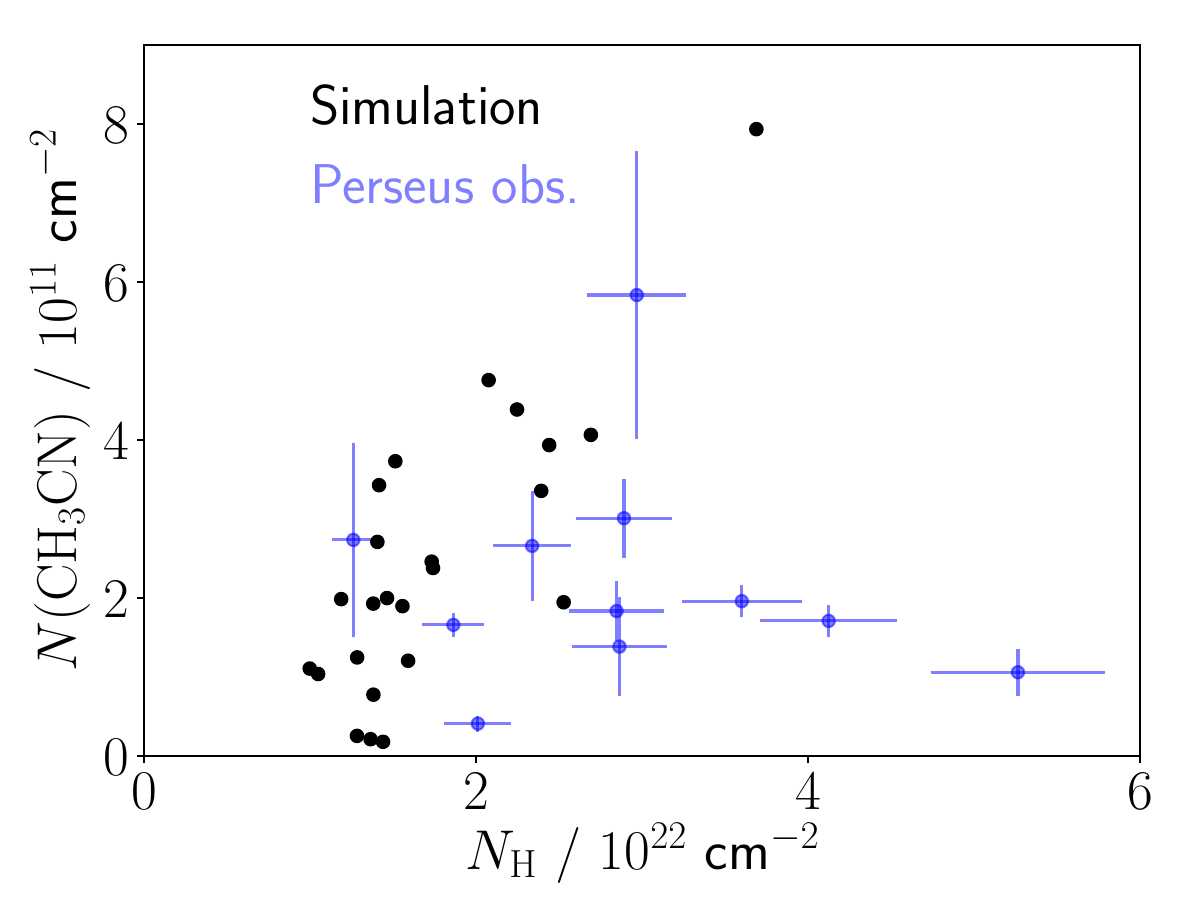}
  \caption{{\it Left:} Gas-phase CH$_3$OH versus total column density for cores from the simulation (black circles), and observed cores in Taurus (green circles; \citealt{scibelli2020}) and Perseus (blue circles; \citealt{scibelli2024}). {\it Centre:} Ice-phase CH$_3$OH versus total column density for simulated cores (black circles). The red dashed lines show the observed range of ice-phase methanol abundances in molecular clouds derived from background stars \citep{boogert2015}. {\it Right:} Gas-phase CH$_3$CN versus total column density for cores from the simulation (black circles), and observed cores in Perseus (blue circles; \citealt{scibelli2024}).}
  \label{fig:coreobs}
\end{figure*}

We perform magnetohydrodynamical simulations using the {\sc arepo} moving-mesh code \citep{springel2010,pakmor2011}, with the addition of a comprehensive suite of physical processes to model the thermodynamical evolution of the gas and dust \citep{glover2007,glover2012,clark2012b}. This includes a simplified chemical network for the formation of H$_2$ and CO, based on \citet{gong2017} with some changes as described in \citet{hunter2023}, and a self-consistent treatment of shielding from the background ultraviolet (UV) radiation field \citep{clark2012}. Sink particles, representing newly-formed stars or stellar systems \citep{bate1995}, are introduced with a threshold density of $2 \times 10^{-16} \gcc$ and a formation radius of $9 \times 10^{-4} \pc$, following the criteria for creation from \citet{tress2020}.

Our simulation consists of two spherical, uniform-density $10^4 \msun$ clouds of radius $19 \pc$, for an initial number density of hydrogen nuclei $\nh = 10 \pcc$, with their centres separated by $38 \pc$ so the clouds are initially just touching. The clouds are given a turbulent velocity field with velocity dispersion $0.95 \kms$, bulk motions of $7 \kms$ toward each other, and a magnetic field of $3 \ug$ along the collision axis. Initial gas and dust temperatures are set to $300 \kel$ and $15 \kel$ respectively. The background UV radiation field is set to $1.7$ times the \citet{habing1968} value, the cosmic ray ionisation rate to $10^{-16} \, {\rm s^{-1}}$ per H atom (consistent with estimates for the diffuse ISM and molecular clouds; \citealt{indriolo2015,sabatini2023}), the dust-to-gas ratio to $0.01$, and the carbon and oxygen abundances to the values in Table \ref{tab:abun}, taken from \citet{sembach2000}. The `metal' abundance, representing heavier elements such as silicon, is set to $10^{-7}$. The simulation is run for $5.53 \myr$, by which point $102 \msun$ of material ($0.5 \%$ of the intiial mass) has been converted into sink particles.

We use Monte Carlo tracer particles \citep{genel2013} to follow the evolution of individual parcels of gas through the simulation, recording the properties of their parent cells at intervals of $44 \kyr$. We select $10^5$ of these particles with final densities in the range $10^2-10^6 \pcc$, and use their evolutionary histories to calculate the chemical composition with the NEATH framework\footnote{https://fpriestley.github.io/neath/} (Non-Equilibrium Abundances Treated Holistically; \citealt{priestley2023a}). This involves using the time-dependent gas-grain code {\sc uclchem} \citep{holdship2017} to evolve the UMIST12 \citep{mcelroy2013} reaction network, with some modifications to ensure the results are consistent with the internal {\sc arepo} abundances of H$_2$ and CO. Adopted elemental abundances are listed in Table \ref{tab:abun}; these are also taken from \citet{sembach2000}, with silicon and magnesium reduced by a factor of 100 to represent their incorporation into refractory dust grains in the denser ISM \citep{jenkins2009}.

In addition to the gas-phase reactions, {\sc uclchem} also models grain surface chemistry by tracking the freeze-out of molecules onto grain surfaces and their subsequent desorption back into the gas phase. Depletion rates are calculated following \citet{rawlings1992}, and desorption via H$_2$ formation, cosmic ray impacts, and the UV field (both background and cosmic ray-generated) is modelled using the approach of \citet{roberts2007}. {We assume desorption efficiencies of $0.01$ molecule per H$_2$ formation, $10^5$ per cosmic ray impact, and $0.1$ per UV photon, as recommended in \citet{roberts2007}.} The reaction network we use here does not explicitly include reactions between ice mantle species, but allows for some fraction of depleted molecules to be hydrogenated instantaneously upon incorporation into the ice phase.

In this paper, we focus on two of the simplest COMs: methanol (CH$_3$OH) and methyl cyanide (CH$_3$CN). As well as representing the broader O- and N-bearing families of COMs, the two molecules have distinct formation pathways. In our chemical network, methanol is formed by the hydrogenation of CO on grain surfaces, which we assume happens with an efficiency of $f_{\rm CH_3OH} = 0.1$ - {two separate CO freeze-out reactions are included in the network, the products being either CO or CH$_3$OH ice, with branching ratios ($0.9$ and $0.1$ respectively) modifying the base rate for CO depletion. Ice-phase CH$_3$OH participates in no reactions other than desorption processes, while gas-phase CH$_3$OH undergoes two-body reactions, photodissociation and freeze-out in an identical manner to other species in the network.} We find that this value {of $f_{\rm CH_3OH}$} produces a good agreement with observed methanol abundances in cores, but our main conclusions hold regardless of the precise value chosen (see Figure \ref{fig:ratelow}). {The impact of the underlying assumptions of this model, such as instantaneous hydrogenation, are discussed in Section \ref{sec:discussion}.} Methyl cyanide is {only formed via gas-phase reactions, primarily} by the reaction between CH$_3^+$ and HCN to form CH$_3$CNH$^+$ followed by recombination with free electrons. We adopt the dissociative recombination reaction rate recommended by \citet{giani2023} rather than the UMIST12 value; we discuss this choice in Appendix \ref{sec:g23}.

\section{Results}

\begin{figure*}
  \centering
  \includegraphics[width=0.32\textwidth]{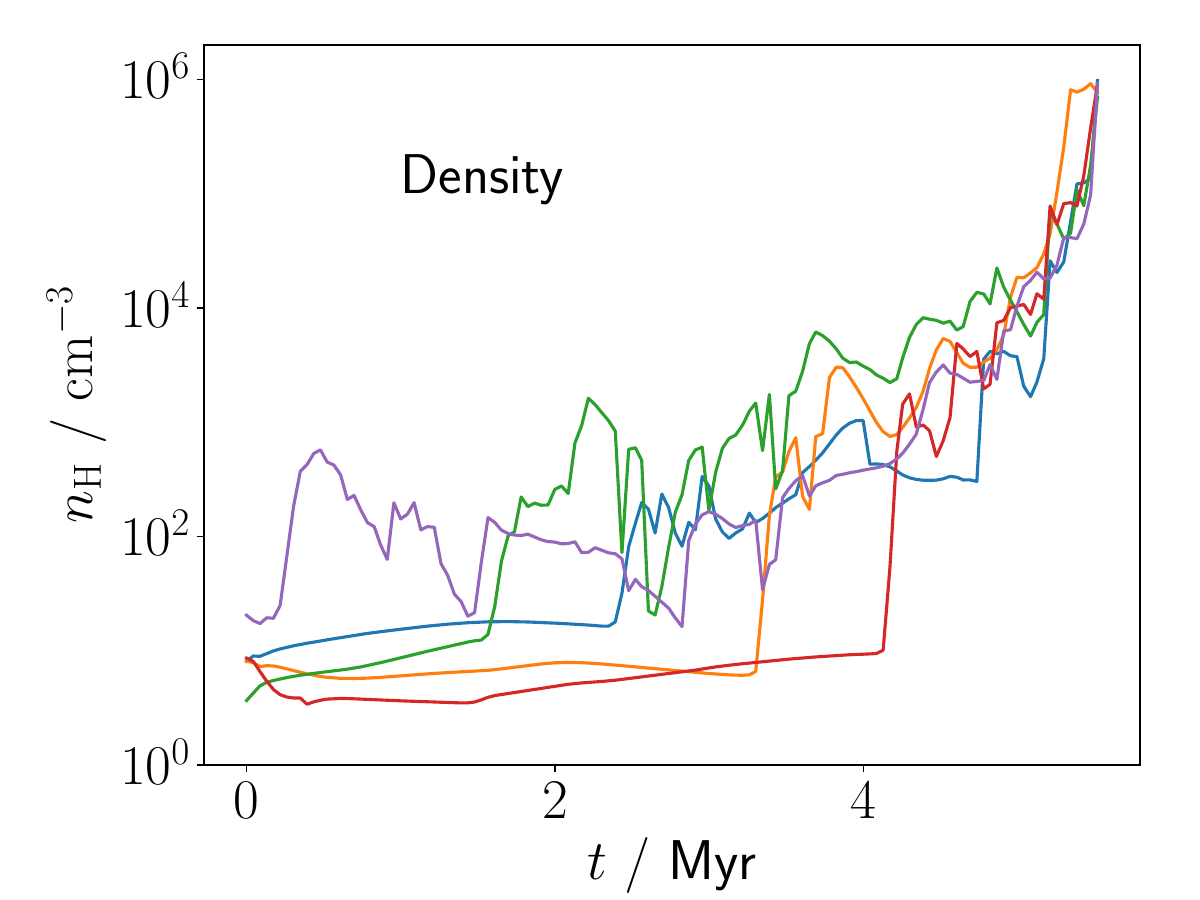}
  \includegraphics[width=0.32\textwidth]{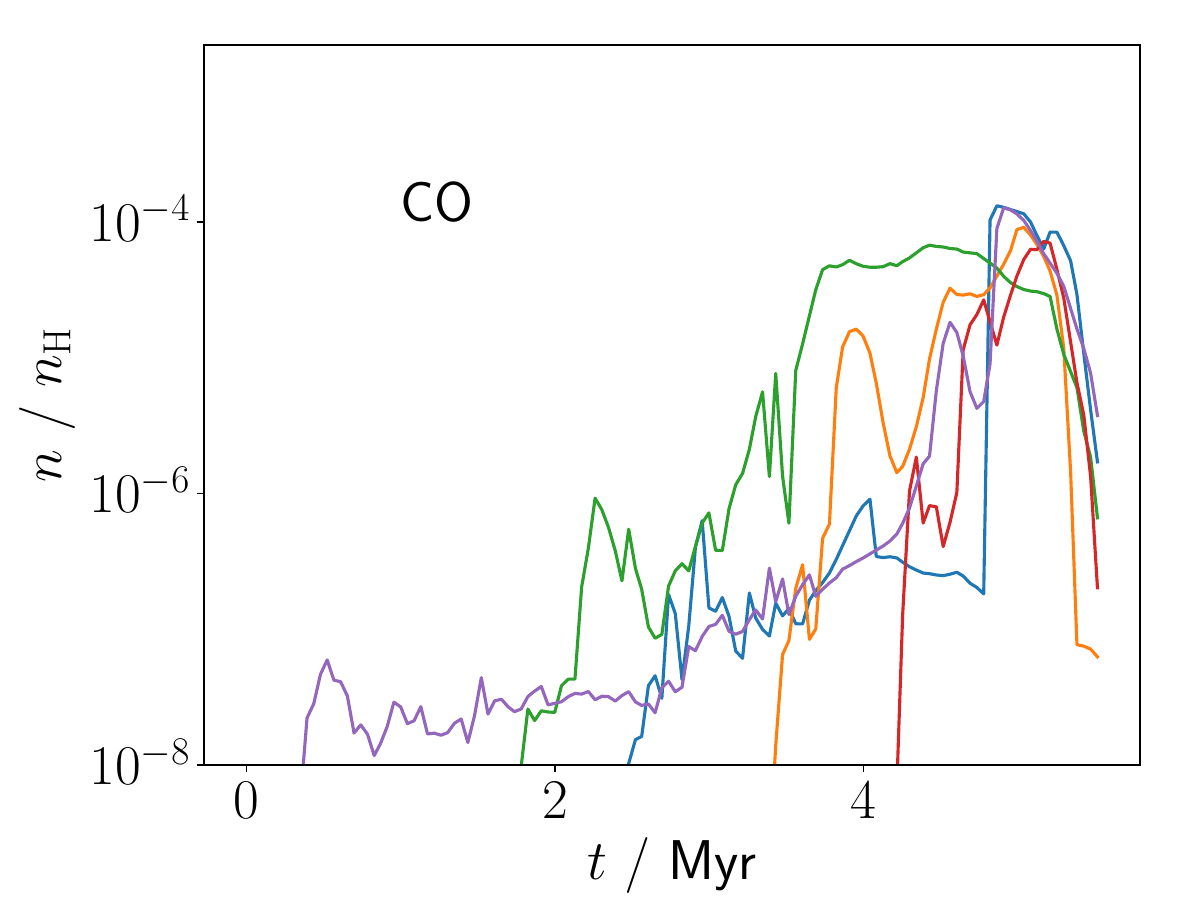}
  \includegraphics[width=0.32\textwidth]{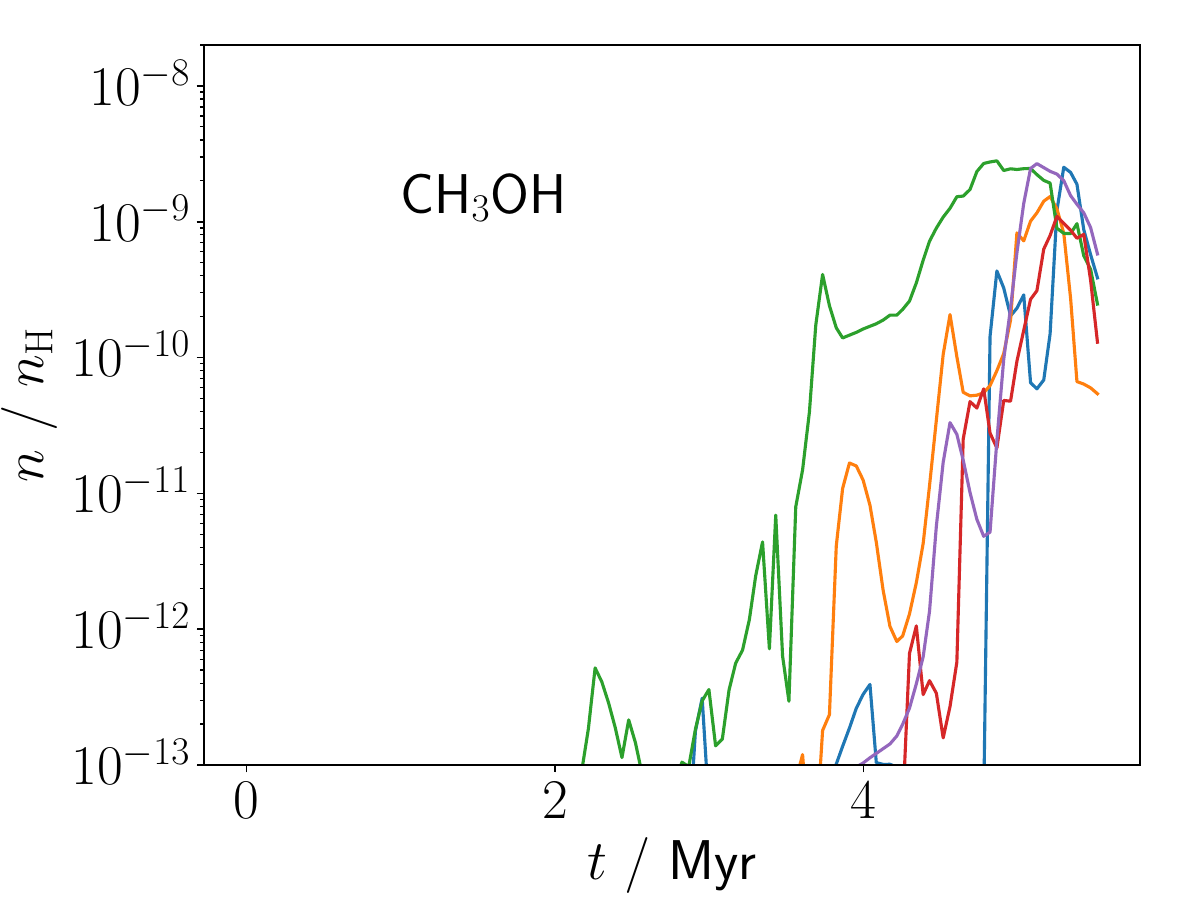}
  \caption{Time evolution of the gas density (left) and CO (centre) and CH$_3$OH (right) abundances for five representative tracer particles (coloured lines), all with final densities of $\sim 10^6 \pcc$.}
  \label{fig:timeevol}
\end{figure*}

\begin{figure*}
  \centering
  \includegraphics[width=0.32\textwidth]{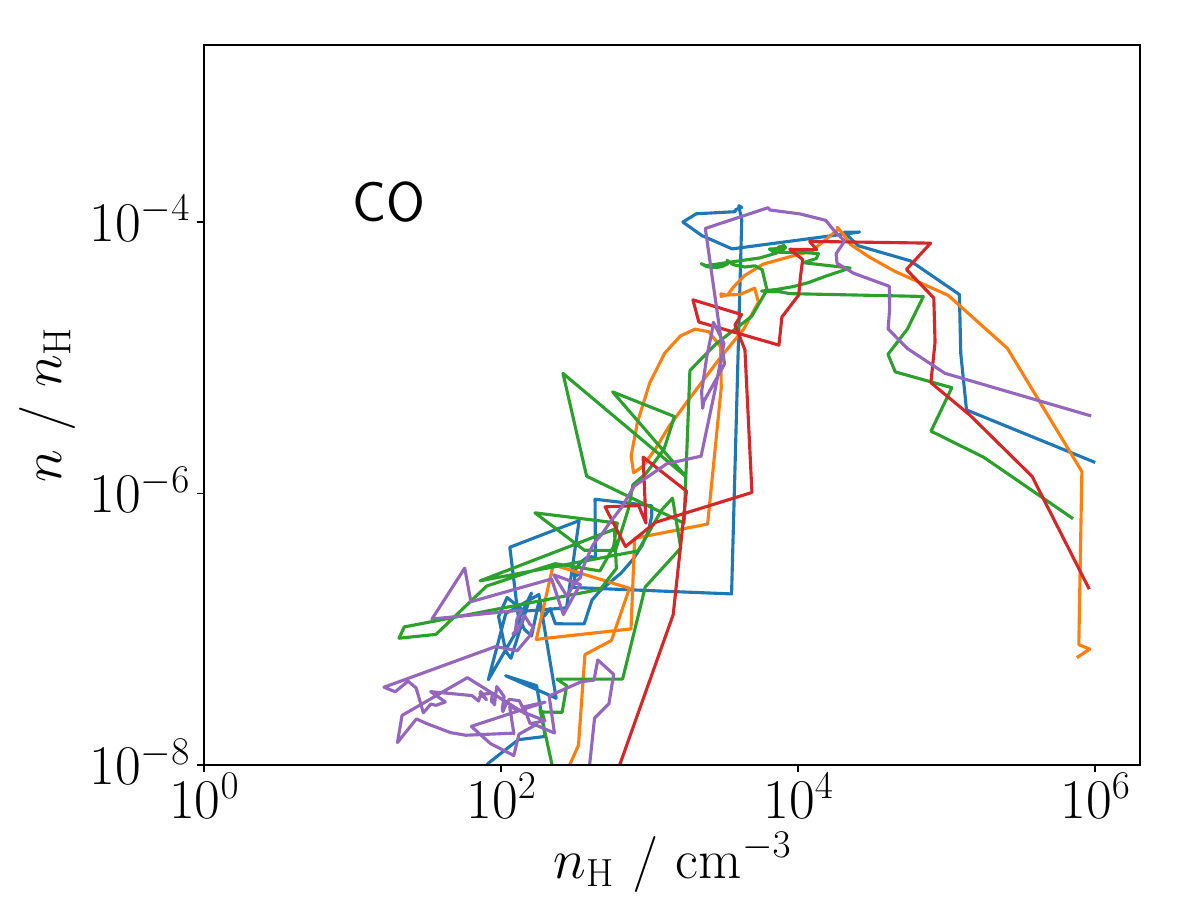}
  \includegraphics[width=0.32\textwidth]{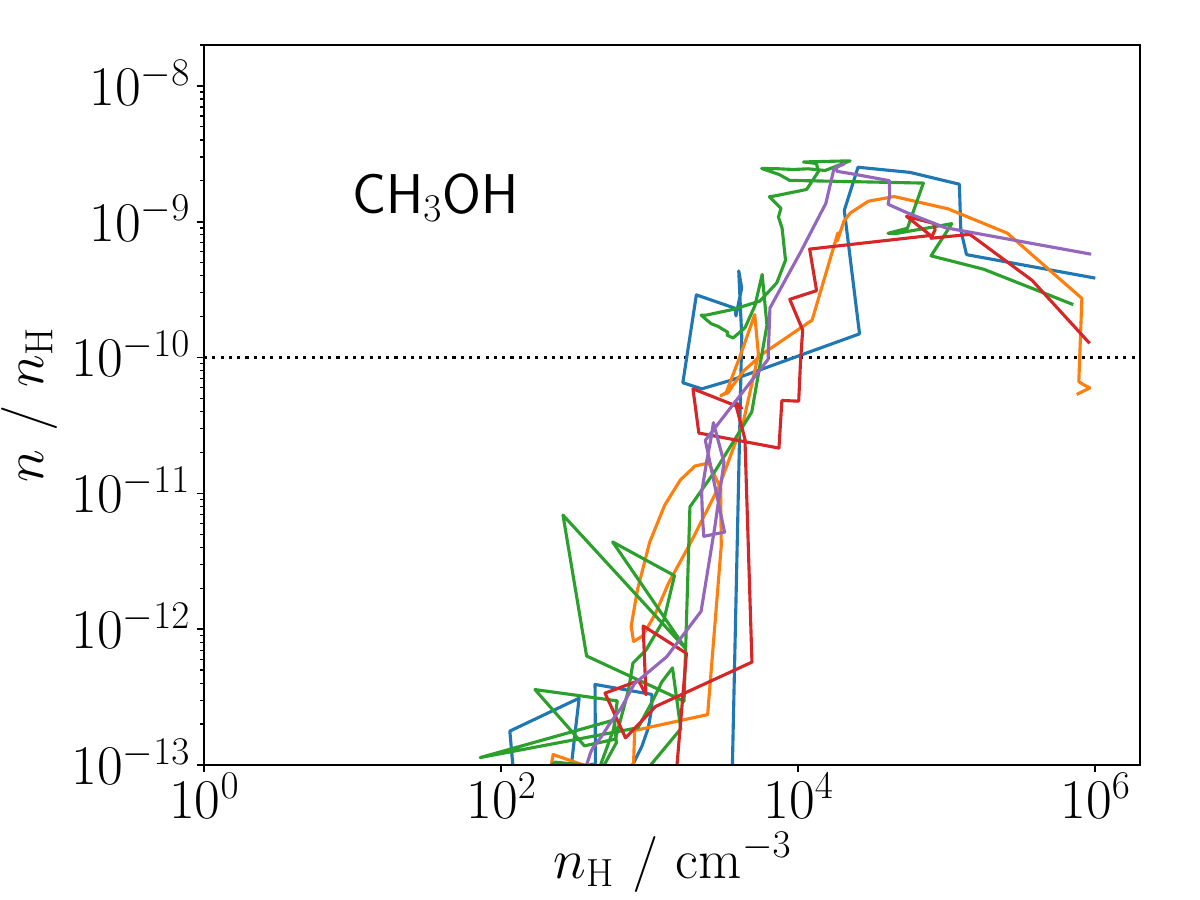}
  \includegraphics[width=0.32\textwidth]{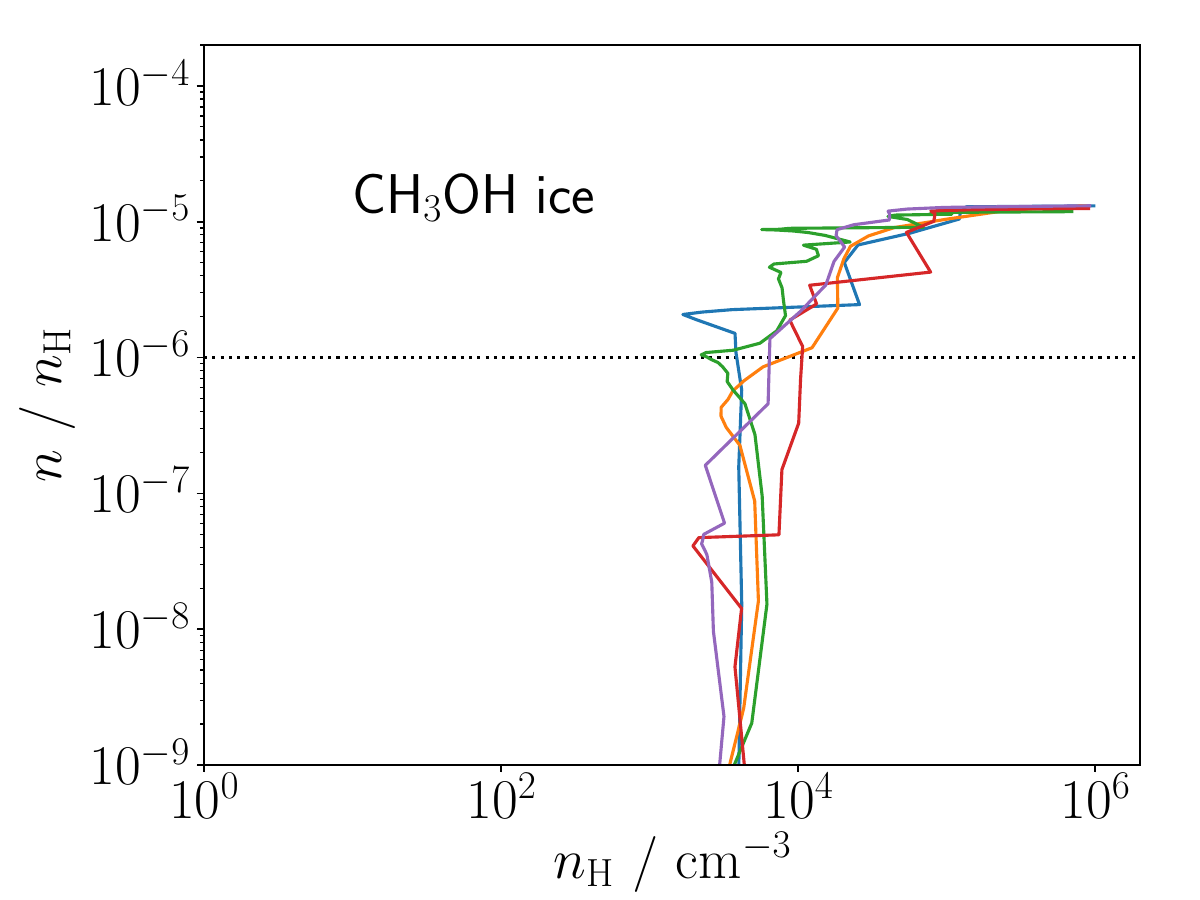}
  \caption{Evolution of gas-phase CO (left), gas-phase CH$_3$OH (centre) and ice-phase CH$_3$OH (right) abundances with gas density for the same five tracer particles as in Figure \ref{fig:timeevol} (coloured lines). Dashed horizontal lines show the threshold abundances used for defining the formation densities in Figure \ref{fig:formdens}.}
  \label{fig:densevol}
\end{figure*}

\begin{figure*}
  \centering
  \includegraphics[width=0.32\textwidth]{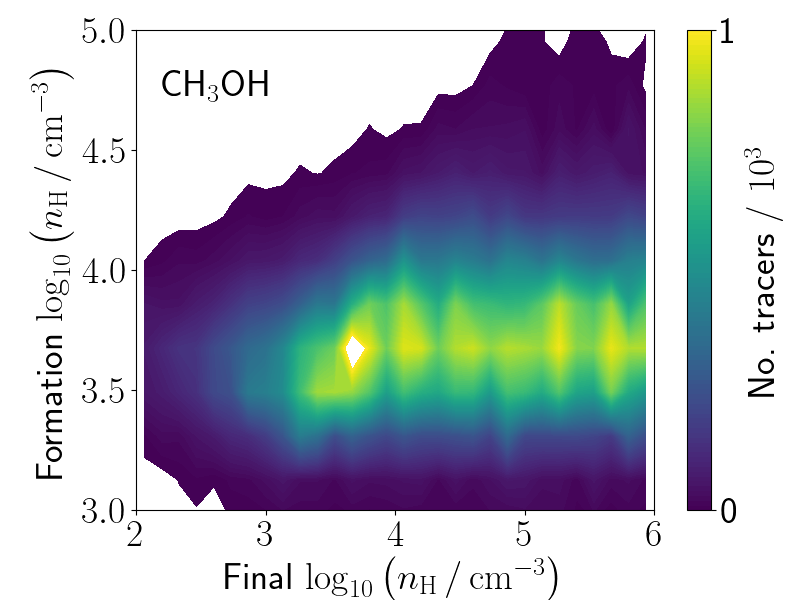}
  \includegraphics[width=0.32\textwidth]{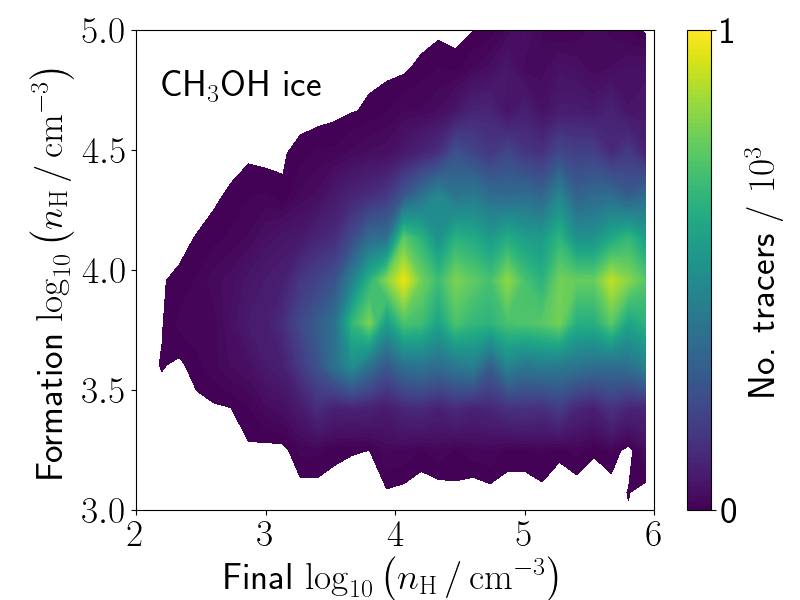}
  \includegraphics[width=0.32\textwidth]{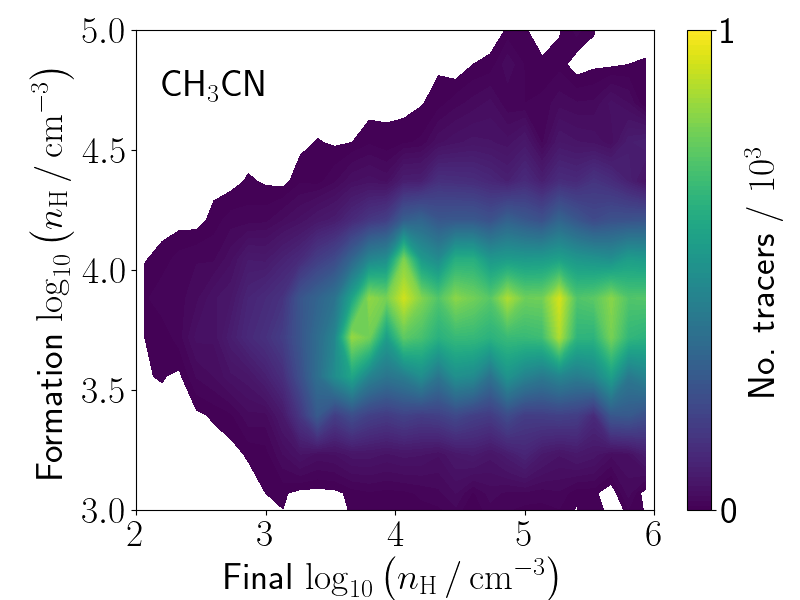}
  \caption{Distribution of tracer particles by final density and formation density of gas-phase CH$_3$OH (left), ice-phase CH$_3$OH (centre), and gas-phase CH$_3$CN (right), for threshold abundances of $10^{-10}$ for gas-phase CH$_3$OH, $10^{-6}$ for ice-phase CH$_3$OH, and $10^{-12}$ for gas-phase CH$_3$CN.}
  \label{fig:formdens}
\end{figure*}

\begin{figure*}
  \centering
  \includegraphics[width=0.32\textwidth]{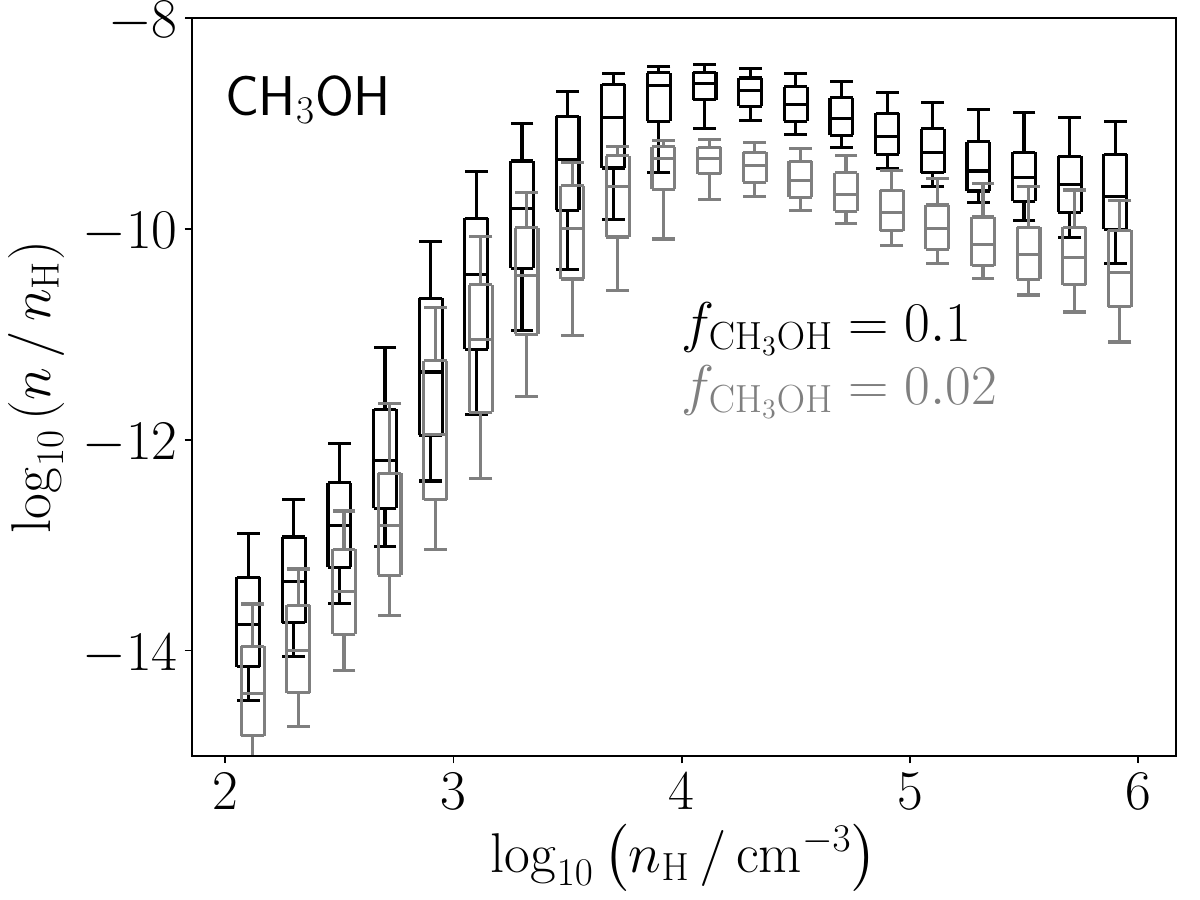}
  \includegraphics[width=0.32\textwidth]{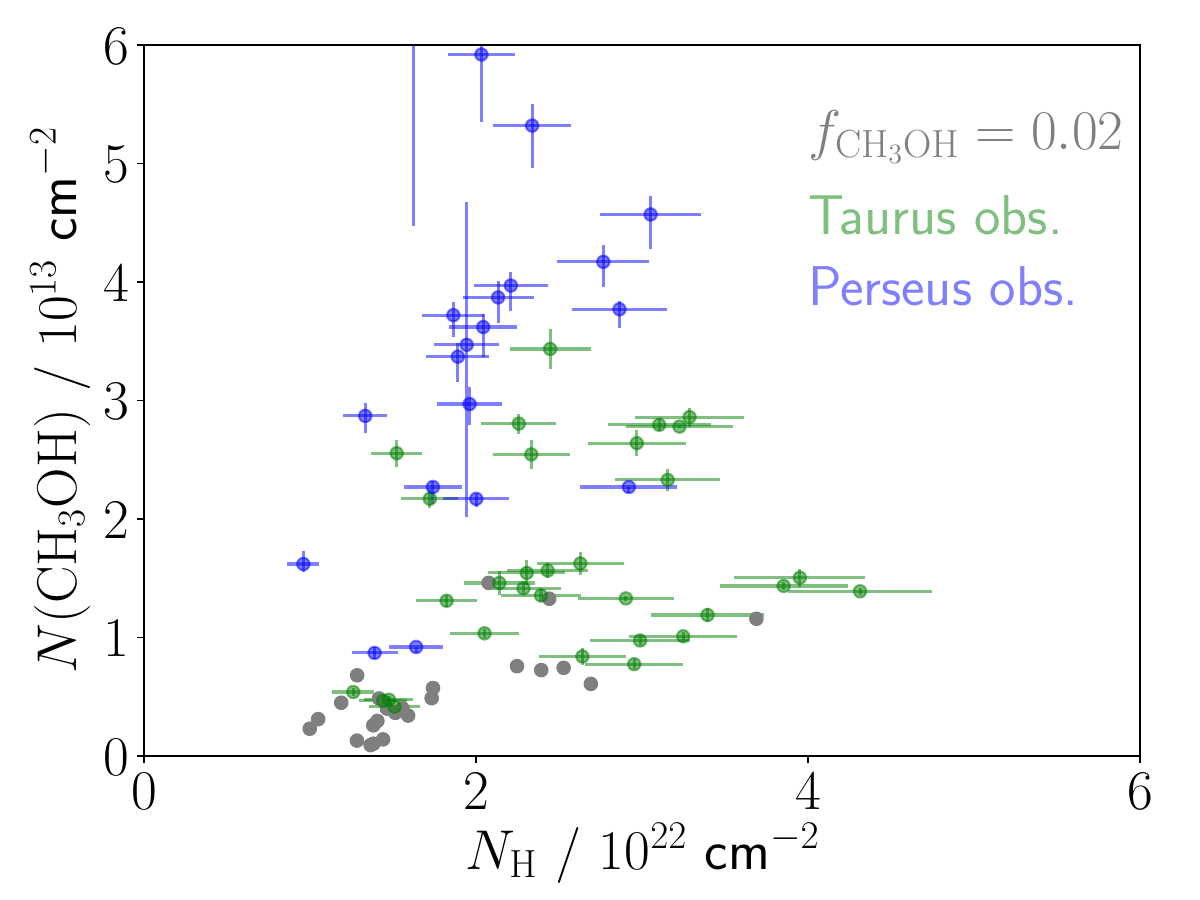}
  \includegraphics[width=0.32\textwidth]{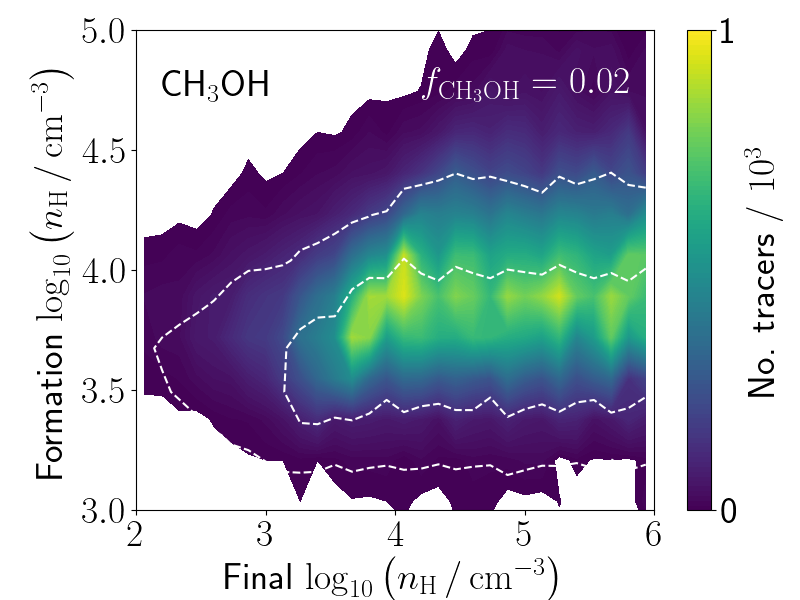}
  \caption{{\it Left:} Gas-phase CH$_3$OH abundance versus density for chemical models with $f_{\rm CH_3OH} = 0.1$ (black) and $0.02$ (grey). Boxes show median values and 25th/75th percentiles, whiskers the 10th/90th percentiles. {\it Centre:} Gas-phase CH$_3$OH versus total column density for cores from the simulation with $f_{\rm CH_3OH} = 0.02$ (grey circles), and observed cores in Taurus (green circles; \citealt{scibelli2020}) and Perseus (blue circles; \citealt{scibelli2024}). {\it Right:} Distribution of tracer particles by final density and formation density of gas-phase CH$_3$OH with $f_{\rm CH_3OH} = 0.02$, for a threshold abundance of $10^{-10}$. {Dashed white lines show the $100$ and $500$-tracer contours from Figure \ref{fig:formdens}.}}
  \label{fig:ratelow}
\end{figure*}

Figure \ref{fig:coldens} shows column density maps of total hydrogen nuclei ($N_{\rm H}$), and of gas- and ice-phase CO, CH$_3$OH and CH$_3$CN after $5.53 \myr$, by which point the initially-atomic clouds have formed a substantial mass of denser molecular material at the collision interface. Much of the diffuse structure seen in the total column map is not visible in any molecular tracer, with CO being restricted to the denser, better-shielded regions of the cloud. CH$_3$OH and CH$_3$CN are even more limited in terms of the fraction of cloud area with significant column densities. However, they are not entirely restricted to the peaks in the column density distribution; substantial gas-phase columns of both COMs exist along moderate-density sightlines through the cloud. This is also true for the ice-phase column densities which are again extended beyond the $0.1 \pc$-scale core structures, although to a lesser extent for CH$_3$CN ice, which requires particularly high densities to both form in the gas-phase and then deplete onto grain surfaces.

Figure \ref{fig:volabun} shows the distribution of CO, CH$_3$OH and CH$_3$CN abundances in both gas and ice phases as a function of volume density. In all cases, abundances rise sharply to a peak value at densities between $10^{3-4} \pcc$, beyond which point gas-phase abundances begin to decline due to freeze-out, while ice-phase abundances remain relatively constant. This decline is less pronounced for CH$_3$OH, where the depletion of the gas-phase molecule is partially balanced by its formation in the mantle via CO depletion and the subsequent desorption of the CH$_3$OH ice.

We also show results for when the physical evolution of the system is ignored: {tracer evolutionary histories are modified so that the density, temperature, and shielding columns are equal to their final values for the entire $5.53 \myr$ duration of the simulation, and the chemical model rerun on these non-evolving histories from the same (atomic) initial conditions as the dynamic model. The approach of evolving the chemistry with fixed physical properties} is common in models of COM formation in cores \citep{vasyunin2017,riedel2023}. The two approaches produce broadly consistent results up to densities of $\sim 10^4 \pcc$, as the chemical timescales are comparable to or shorter than the dynamical timescales at these densities. The abundances in the dynamic model are therefore quite close to their equilibrium values for densities below $10^4 \pcc$ \citep{holdship2022,priestley2023a,rawlings2024}, making the use of static chemical models justified in this regime.

At higher densities, however, the static approach significantly underestimates the gas-phase abundances of all species due to allowing an unrealistic degree of freeze-out; the actual amount of time spent at these densities is a tiny fraction of the $\sim \myr$ timescales over which static chemical models are typically evolved \citep{priestley2023a}. CH$_3$OH, with its ice-phase formation pathway, is less affected than CO and CH$_3$CN, as the excess depletion is partially balanced by excess formation via CO freeze-out, but its gas-phase abundance in the static model can still be an order of magnitude lower than the correct value.

In static chemical models of cores, most of the COM content exists in the central, well-shielded regions, where densities are significantly higher than $10^4 \pcc$ \citep{vasyunin2017,riedel2023}. These models are therefore likely to systematically underestimate the gas-phase abundances of COMs, compared to models with a more realistic treatment of core formation and evolution. {Molecular column densities, integrated through the entire cloud, do not vary by more than a factor of $2-3$ between static and dynamic models, because the fraction of mass above $10^4 \pcc$ along the line-of-sight is rarely high enough for the large abundance differences in this regime to become apparent. However, changing the abundances in this high-density material may have a significant impact on predictions for molecular line emission \citep[e.g.][]{yin2021}, particularly for transitions with high critical densities. Chemical models utilising full evolutionary histories should be preferred when applied to radiative transfer calculations.}

To enable comparison with observational results, we create a dendrogram \citep{rosolowsky2008} of the $N_{\rm H}$ map using {\sc astrodendro}\footnote{http://www.dendrograms.org/}, with a threshold of $10^{22} \pcs$, a minimum branching significance of $5 \times 10^{21} \pcs$, and a minimum pixel number of $9$. We identify dendrogram leaves as `cores'; column densities are then calculated by averaging all pixels within $0.1 \pc$ of the peak $N_{\rm H}$ pixel for each leaf, approximating the effects of beam size for single-dish observations of nearby ($\sim 100 \pc$) clouds. We compare the results to values inferred from observational data in Figure \ref{fig:coreobs}.

The simulated cores have comparable gas-phase CH$_3$OH column densities to observed starless and prestellar cores in the Taurus \citep{scibelli2020} and Perseus \citep{scibelli2024} molecular clouds, which suggests that our methanol formation efficiency of $f_{\rm CH_3OH} = 0.1$ is correctly reproducing the typical gas-phase CH$_3$OH abundances in real cores. Ice-phase CH$_3$OH abundances are also comparable to those observed towards background sources along quiescent sightlines \citep{boogert2015}, and gas-phase CH$_3$CN column densities are in good agreement with those of Perseus cores for which this molecule is detected \citep{scibelli2024}. Our simulated cores appear to be similar to real objects in terms of their COM content, which {suggests that} their COM formation histories {may also be} similar; {this is our working assumption for the remainder of the paper, subject to caveats regarding the limitations of our chemical model discussed in Section \ref{sec:discussion}}.

The time evolution of the gas density and the CO and CH$_3$OH abundances for five tracer particles, {randomly selected from those with final densities above $6 \times 10^5 \pcc$,} are shown in Figure \ref{fig:timeevol}. Despite having {similar} final densities of $\sim 10^6 \pcc$, the tracers show very different evolutionary histories, with multiple episodes of compression and rarefaction preceding a more-or-less monotonic increase in density to the final value. The diversity in physical evolution results in similarly-diverse chemical evolution, with tracers forming (and occasionally losing) significant gas-phase abundances of CO and CH$_3$OH at different times. However, we note that the late-time monotonic increase in density results in a rapid decline in the gas-phase CO abundance for all tracers due to freeze-out, and that the production of substantial quantities of gas-phase CH$_3$OH has already occured prior to this {onset of severe CO depletion}.

While the time evolution of the tracer abundances appears quite disparate, the evolution with density\footnote{Note that this is the evolution of abundance with density for individual tracer particles over the course of the simulation, as opposed to the distribution of abundance versus density at $5.53 \myr$ for {\it all} tracers in Figure \ref{fig:volabun}.}, shown in Figure \ref{fig:densevol}, is much more uniform. The CO abundance begins to sharply increase as the gas density approaches $10^3 \pcc$, and reaches a peak value of $\sim 10^{-4}$ at a density of around $10^4 \pcc$. This is mirrored by the gas-phase abundance of CH$_3$OH, which also begins rising sharply at a density of $10^3 \pcc$, and has essentially reached its peak value by $10^4 \pcc$ for all five tracers. The abundance of CH$_3$OH ice rises even more sharply with density; the trajectories in density-abundance space are close to vertical at a density of around $3 \times 10^3 \pcc$.

The rapid increase in molecular abundances once a certain density threshold has been reached allows us to define a `formation density' for each species, characterising the point at which significant quantities of the molecule (relative to its peak abundance) begin to appear. We identify this as the density at which the abundance of a species first crosses a threshold chosen to be around $10 \%$ of the maximum in Figure \ref{fig:volabun}, although due to the steepness of the trajectories in Figure \ref{fig:densevol}, our conclusions are relatively insensitive to the exact value chosen; increasing the CH$_3$OH threshold by an order of magnitude only results in a factor of 2 difference in the typical formation densities. Figure \ref{fig:formdens} shows the distribution of tracers by final density at $5.53 \myr$ versus the formation density of gas- and ice-phase CH$_3$OH and gas-phase CH$_3$CN, with abundance thresholds of $10^{-10}$, $10^{-6}$ and $10^{-12}$ respectively. All three species form at densities below $10^4 \pcc$, with significant quantities of gas-phase CH$_3$OH being produced at densities of only a few $10^3 \pcc$. The formation density is effectively constant with regard to the final density of the tracers: material which ends up incorporated into dense cores forms its COM content at the same early evolutionary phase as the surrounding diffuse cloud material.

The methanol formation efficiency of $f_{\rm CH_3OH} = 0.1$ used above results in good agreement with its observed gas- and ice-phase abundances (Figure \ref{fig:coreobs}), but we show in Figure \ref{fig:ratelow} that our conclusions regarding the formation of COMs are not affected by the exact choice of this parameter. Reducing the formation efficiency by a factor of five results, unsurprisingly, in a lower CH$_3$OH abundance at all densities, and gas-phase column densities which are now in some tension with the observed values. However, we still find formation densities below $10^4 \pcc$, even for an unchanged abundance threshold of $10^{-10}$ which is close to the maximum value reached. Reducing the threshold to account for the lower peak abundance leads to almost indistinguishable results from those in Figure \ref{fig:formdens}.

\section{Discussion}

\subsection{Limitations of the chemical model}
\label{sec:discussion}

{In addition to the efficiency parameter $f_{\rm CH_3OH}$, our treatment of CH$_3$OH formation on grain surfaces relies on the assumptions that hydrogenation is instantaneous (for some fraction of adsorbed CO molecules), that $f_{\rm CH_3OH}$ has a fixed value, and that no further grain surface processes affect the CH$_3$OH ice after formation. The first of these assumptions is reasonable: simulations which explicitly follow hydrogenation of CO in ice mantles \citep[e.g.][]{garrod2022,molpeceres2024} find that the delay between adsorption of CO and formation of CH$_3$OH ice is a few kyr, effectively instantaneous in the context of our simulations with a chemical timestep of $44 \kyr$. The assumption of a fixed $f_{\rm CH_3OH}$ may not hold in all environments, as adsorbed CO may preferentially be converted into CO$_2$ rather than CH$_3$OH in unshielded environments with dust temperatures $\gtrsim 12 \kel$ \citep{garrod2011}. Typical grain temperatures in our simulation are $\sim 11 \kel$ for a density of $3000 \pcc$ (around the onset of CH$_3$OH formation) and $10 \kel$ at $10^4 \pcc$, so this effect would at most push the formation density of CH$_3$OH up to the latter value, leaving our conclusions largely unchanged. Finally, additional processes not considered here, such as hydrogen abstraction reactions and photodissociation of mantle species, could reduce the abundance of CH$_3$OH in the ice and, subsequently, gas phases. As these processes only operate on preexisting CH$_3$OH ice, our conclusions regarding the onset of its formation would again be unaffected by their inclusion.}

{Our treatment of mantle desorption processes, following \citet{roberts2007}, assumes that species are returned to the gas phase with a fixed efficiency per event (H$_2$ formation or cosmic ray/UV photon impact). The appropriate values for these efficiencies are uncertain, and the approach ignores possible effects such as localised energy deposition \citep{pantaleone2021} and desorption of fragments rather than parent species \citep{bertin2016}, which would tend to reduce the abundance of gas-phase CH$_3$OH. We again note that this would not change our conclusions regarding early COM formation. Figures \ref{fig:ratelow} and \ref{fig:g23} demonstrate that the absolute value of a species' abundance has little effect on its formation density, as altering reaction rates leaves the {\it relative} behaviour of the abundance with density mostly unchanged. In any case, the formation of CH$_3$OH ice would be almost entirely unaffected by changes to desorption processes, given the extremely low gas/ice abundance ratio of this molecule, while desorption has a negligible effect on the gas-phase CH$_3$CN abundance except at the very highest densities, well beyond the density of its formation via gas-phase reactions.}

{Our conclusions are robust with respect to changes in the chemical model because the early formation of COMs is primarily a physical effect: in realistic simulations of turbulent molecular clouds, material spends significant periods of time at moderate densities (Figure \ref{fig:timeevol}) before becoming gravitationally bound. This allows the freeze-out of around $10 \%$ of the total elemental carbon budget in the form of CO ice at gas densities of a few $10^3 \pcc$ (Figure \ref{fig:volabun}). Any chemical model which forms CH$_3$OH via hydrogenation of CO ice will therefore reach the same conclusions regarding the early formation of this molecule. The prolonged evolution at moderate densities also allows species with gas-phase formation pathways, such as CH$_3$CN, to exist in quantities which would not be anticipated by studies relying on simplified physical models. \citet{garrod2022} have also argued for early formation of COMs, but in their case this is due to the inclusion of UV-induced photochemical reactions in ice mantles. Their physical model is a free-fall collapse from an initial density of $3000 \pcc$, with pristine atomic chemical initial conditions; we argue instead that by the time material reaches this density, it should already have formed a significant fraction of its final COM content.}

\subsection{Implications}

While there is significant ambiguity in defining what counts as a `core' and in distinguishing between `prestellar' and `starless' objects, both observationally and in simulations \citep[e.g.][]{offner2022,scibelli2023}, a volume density threshold of around $10^4 \pcc$ is a reasonable choice for separating core material from the ambient cloud \citep{bergin2007}. \citet{lada2010} argue that gas above this density is directly associated with star formation activity, whereas lower-density material is not. Similarly, we find in our simulations that material crossing this density threshold almost never returns to lower densities \citep{priestley2023b}, which can be seen in the nearly-monotonic increase in tracer densities in Figure \ref{fig:timeevol} once a value of $10^4 \pcc$ has been reached. Under this definition, cores have formed virtually all their COM content before they actually become cores, with COM formation starting at densities of a few $10^3 \pcc$ and being essentially complete by the time the gas reaches a few $10^4 \pcc$.

This stands in stark contrast to the typical approach when modelling COM chemistry in cores \citep[e.g.][]{vasyunin2017,riedel2023}, where the initial abundances are atomic or diffuse-molecular even up to densities of $10^7 \pcc$. \citet{slavicinska2024} recently found that the CH$_3$OH observed in protostellar cores must have formed in H$_2$O-rich rather than CO-rich ice, contradicting its assumed origin in heavily CO-depleted cores, but entirely consistent with our proposed formation in lower-density cloud material, where most CO is still in the gas phase. We suggest that future work on the chemistry of COMs in cores should account for the fact that these molecules likely already exist in significant quantities before the core itself forms as a distinct object. Moreover, the fact that evolved protostellar systems are observed to continually accrete chemically `fresh' material from larger scales via streamers \citep{pineda2020,valdivia2022,valdivia2023,valdivia2024} makes the composition of this material, and its possible COM content, relevant to studies of disc chemistry and planet formation.

\section{Conclusions}

We have simulated the formation of a molecular cloud, and the cores and stars that it contains, from the diffuse ISM, and followed the chemical evolution of the material from atomic initial conditions up to the formation of COMs using a time-dependent gas-grain reaction network. Our approach successfully reproduces the observed abundances of CH$_3$OH and CH$_3$CN in starless and prestellar cores. We find that the onset of COM formation occurs at relatively low densities of a few $10^3 \pcc$, for both gas- and ice-phase formation pathways.

{Although our chemical model is somewhat simplified, particularly in its treatment of grain surface chemistry, this early COM formation is} essentially a direct result of the dynamical evolution of the gas, which can spend a considerable amount of time at these moderate densities before undergoing runaway gravitational collapse. {It therefore seems likely to be replicated in more sophisticated chemical models, although further studies are required to confirm whether this is in fact the case.}

{If COMs do form at the low densities that we find here,} cores {will already be enriched} in COMs from the moment they begin to appear as identifiable objects, {and complex organic chemistry would originate} early, on the scale of molecular clouds, rather than {in the later pre- and protostellar evolutionary phases}. Astrochemical models {would need to} account for this prior {chemical enrichment} to fully understand the delivery of COMs to forming planetary systems.

\section*{Acknowledgements}

FDP, PCC, SER and OF acknowledge the support of a consolidated grant (ST/W000830/1) from the UK Science and Technology Facilities Council (STFC). SS acknowledges the National Radio Astronomy Observatory is a facility of the National Science Foundation operated under cooperative agreement by Associated Universities, Inc. SCOG and RSK acknowledge funding from the European Research Council (ERC) via the ERC Synergy Grant “ECOGAL-Understanding our Galactic ecosystem: From the disk of the Milky Way to the formation sites of stars and planets” (project ID 855130), from the Heidelberg Cluster of Excellence (EXC 2181 - 390900948) “STRUCTURES: A unifying approach to emergent phenomena in the physical world, mathematics, and complex data”, funded by the German Excellence Strategy, and from the German Ministry for Economic Affairs and Climate Action in project ``MAINN'' (funding ID 50OO2206). The team in Heidelberg also thanks for computing resources provided by {\em The L\"{a}nd} through bwHPC and DFG through grant INST 35/1134-1 FUGG and for data storage at SDS@hd through grant INST 35/1314-1 FUGG. RSK also thanks the 2024/25 Class of Radcliffe Fellows for their company and for highly interesting and stimulating discussions. LRP acknowledges support from the Irish Research Council Laureate programme under grant number IRCLA/2022/1165. This research was undertaken using the supercomputing facilities at Cardiff University operated by Advanced Research Computing at Cardiff (ARCCA) on behalf of the Cardiff Supercomputing Facility and the Supercomputing Wales (SCW) project. We acknowledge the support of the latter, which is part-funded by the European Regional Development Fund (ERDF) via the Welsh Government. This research made use of {\sc astrodendro}, a Python package to compute dendrograms of astronomical data (http://www.dendrograms.org/).

\section*{Data Availability}
The data underlying this article will be shared on request.

\bibliographystyle{mnras}
\bibliography{comabun}

\appendix

\section{The CH$_3$CN formation rate}
\label{sec:g23}

\begin{figure*}
  \centering
  \includegraphics[width=0.32\textwidth]{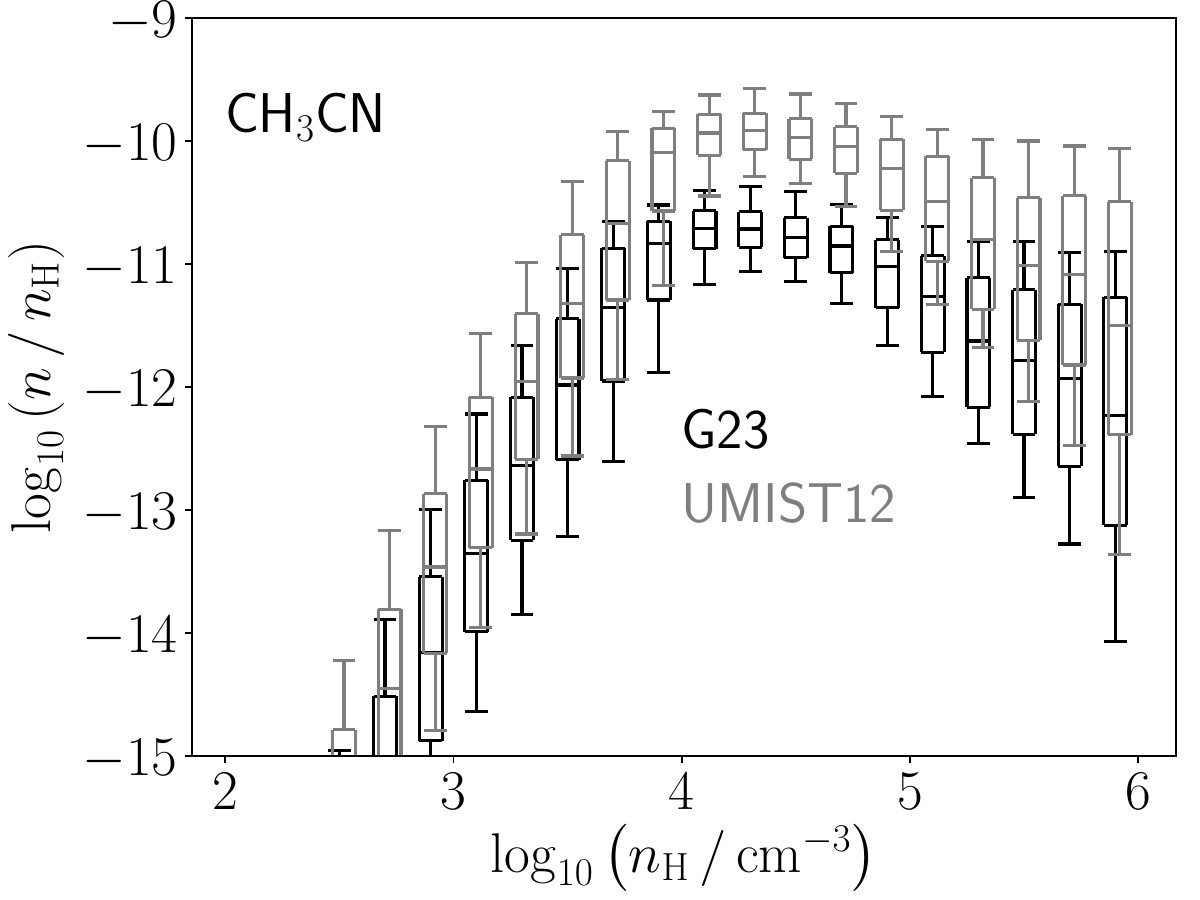}
  \includegraphics[width=0.32\textwidth]{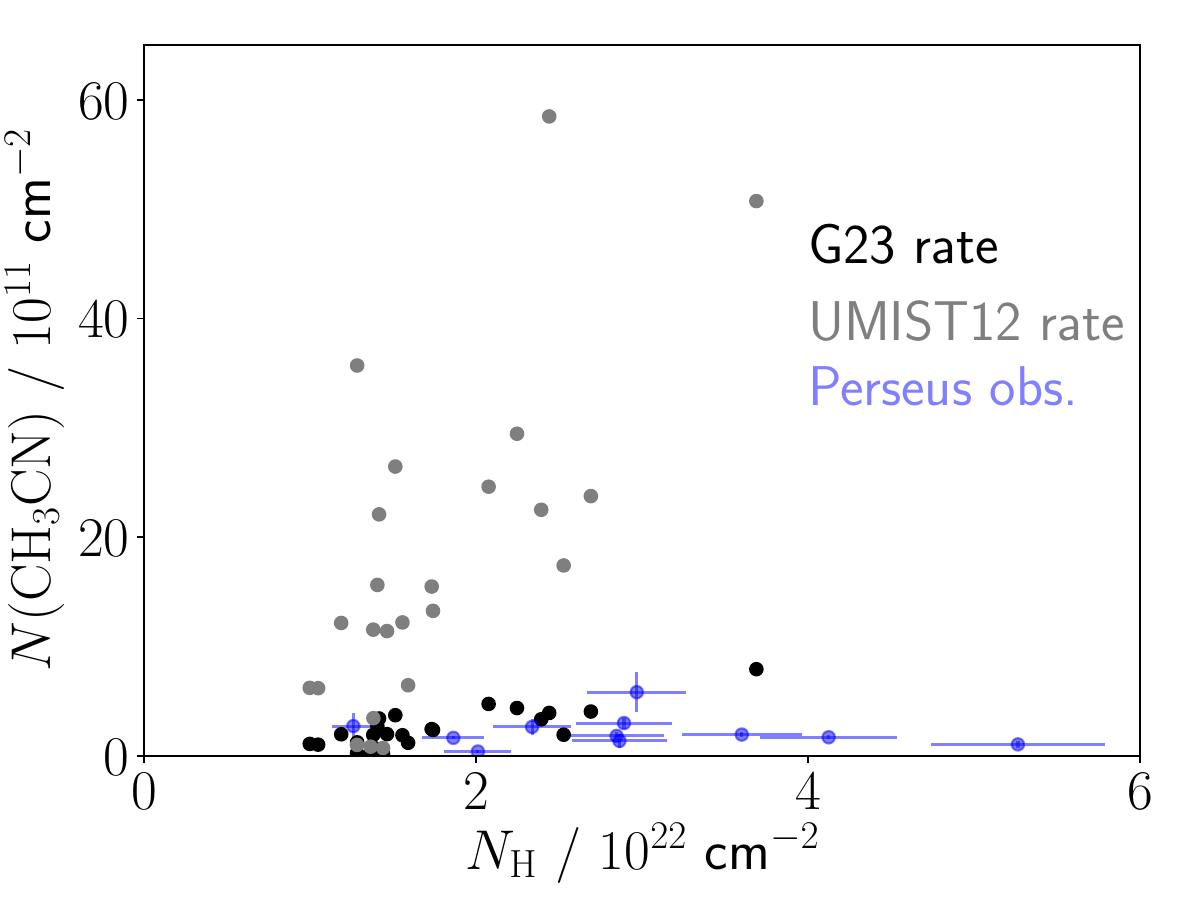}
  \includegraphics[width=0.32\textwidth]{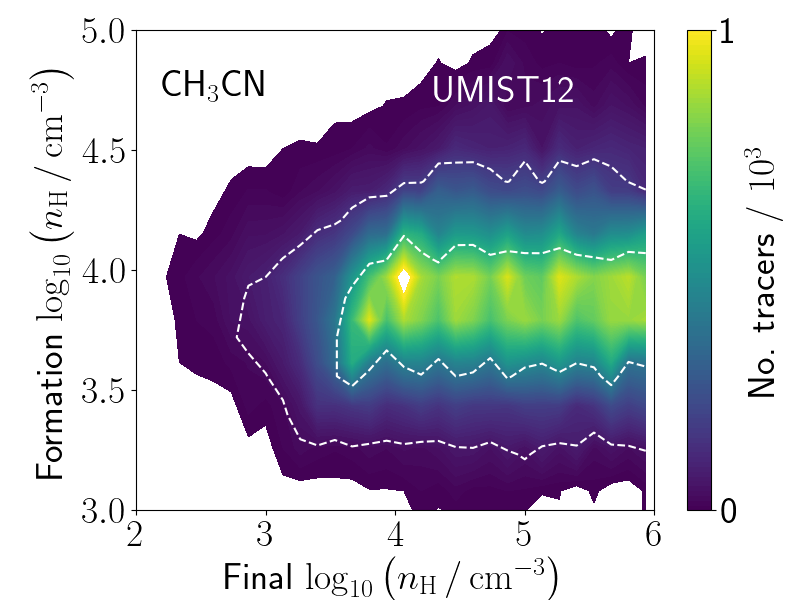}
  \caption{{\it Left:} Gas-phase CH$_3$CN abundance versus density for chemical models with the \citet{giani2023} (black) and UMIST12 (grey) rates for dissociative recombination of CH$_3$CNH$^+$. Boxes show median values and 25th/75th percentiles, whiskers the 10th/90th percentiles. {\it Centre:} Gas-phase CH$_3$CN versus total column density for cores from the simulation using the \citet{giani2023} (black circles) and UMIST12 (grey circles) rates, and observed cores in Perseus (blue circles; \citealt{scibelli2024}). {\it Right:} Distribution of tracer particles by final density and formation density of gas-phase CH$_3$CN using the UMIST12 rate, for a threshold abundance of $10^{-11}$. {Dashed white lines show the $100$ and $500$-tracer contours from Figure \ref{fig:formdens}.}}
  \label{fig:g23}
\end{figure*}

\citet{giani2023} have recently reevaluated the formation pathways of CH$_3$CN, recommending a significantly lower rate for the dissociative recombination reaction between CH$_3$CNH$^+$ and free electrons compared to the UMIST12 value (normalisation at $300 \kel$ of $2.2 \times 10^{-8} \, {\rm cm^3 \, s^{-1}}$ versus $5.3 \times 10^{-7} \, {\rm cm^3 \, s^{-1}}$, in addition to a steeper temperature dependence). We have chosen to adopt their rate coefficients for this reaction, which produces good agreement with observed CH$_3$CN column densities (Figure \ref{fig:coreobs}), but as with the methanol formation efficiency parameter, our main conclusions still hold when using the original UMIST12 values.

Figure \ref{fig:g23} shows that the UMIST12 rate coefficients result in roughly an order of magnitude increase in gas-phase CH$_3$CN abundance at fixed density compared to the \citet{giani2023} values, which leads to a significant tension between predicted and observed column densities for this molecule. However, the qualitative relationship between the abundance and gas density is largely unaffected. We therefore find that the CH$_3$CN formation density, defined using an increased threshold abundance of $10^{-11}$ to reflect the higher peak values, is effectively indistinguishable from the results in Figure \ref{fig:formdens}. Figures \ref{fig:g23} and \ref{fig:ratelow} demonstrate that while the absolute quantity of COMs formed depends on the details of the chemical model, the {\it onset} of COM formation does not, occurring at volume densities consistently below those associated with cores.

\bsp	
\label{lastpage}
\end{document}